\documentclass[fleqn,usenatbib]{mnras}

\usepackage{newtxtext,newtxmath}

\usepackage[T1]{fontenc}

\DeclareRobustCommand{\VAN}[3]{#2}
\let\VANthebibliography\thebibliography
\def\thebibliography{\DeclareRobustCommand{\VAN}[3]{##3}\VANthebibliography}

\usepackage{graphicx}	
\usepackage{amsmath}	
\usepackage{datetime}
\usepackage{caption}
\usepackage{subcaption}
\usepackage{float}
\usepackage{rotating}
\usepackage{pdflscape}
\usepackage{comment}
\usepackage[normalem]{ulem}

\usepackage{xcolor}

\newcommand{\mycomment}[1]{}

\defcitealias{haasSecularTheoryOrbital2011a}{VHSI}

\defcitealias{haasCouplingYoungStellar2011}{VHSII}

\newcommand{\rbar}{\overline{\mathcal{R}}}

\newcommand{\sgra}{SgrA$^\ast$}
\newcommand{\arwv}{\sc{arwv}}
\newcommand{\myr}{\mathrm{Myr}}
\newcommand{\pc}{\mathrm{pc}}

\newcommand{\model}[1]{%
\ifnum#1=1{\sf{}M1}\else
\ifnum#1=2{\sf{}M2}\else
\ifnum#1=3{\sf{}M3}\else
\ifnum#1=4{\sf{}M4}\else
\ifnum#1=5{\sf{}M5}\else
\ifnum#1=6{\sf{}M6}\else
{\tt ???}\fi\fi\fi\fi\fi\fi
}

\title[Dynamical coupling of nearby Keplerian orbits]{Dynamical coupling of Keplerian orbits in a hierarchical four-body system: from the Galactic Centre to compact planetary systems}

\author[M. Singhal, L. Šubr, J. Haas ]{
M. Singhal$^{1}$\thanks{Contact e-mail: \href{mailto:singhal.myank@matfyz.cuni.cz}{singhal.myank@matfyz.cuni.cz}},
L. Šubr$^{1}$,
J. Haas$^{1}$
\\
$^{1}$Astronomical Institute, Faculty of Mathematics and Physics, Charles University, V Holešovičkách 2, 18000 Praha, Czech Republic\\
}

\date{Accepted 2024 May 14. Received 2024 April 16; in original form 2023 November 06}

\pubyear{2024}

\begin{document}
\label{firstpage}
\pagerange{\pageref{firstpage}--\pageref{lastpage}}
\maketitle

\begin{abstract}
This study focuses on the long-term evolution of two bodies in nearby initially coplanar orbits around a central dominant body perturbed by a fourth body on a distant Keplerian orbit. Our previous works that considered this setup enforced circular orbits by adding a spherical potential of extended mass, which dampens Kozai--Lidov oscillations; it led to two qualitatively different modes of the evolution of the nearby orbits. In one scenario, their mutual interaction exceeds the effect of differential precession caused by a perturbing body. This results in a long-term coherent evolution, with nearly coplanar orbits experiencing only small oscillations of inclination. We extend the previous work by (i) considering post-Newtonian corrections to the gravity of the central body, either instead of or in addition to the potential of extended mass, (ii) relaxing the requirement of strictly circular orbits, and (iii) removing the strict requirement of complete Kozai--Lidov damping. Thus, we identify the modes of inter-orbital interaction described for the zero-eccentricity case in the more general situation, which allows for its applicability to a much broader range of astrophysical systems than considered initially. In this work, we scale the systems to the orbits of S-stars; we consider the clockwise disc to represent the perturbing body, with post-Newtonian corrections to the gravity of Sagittarius A* playing the role of damping potential. Considering post-Newtonian corrections, even stellar-mass central bodies in compact planetary systems can allow for the coupled evolution of Keplerian orbits.
\end{abstract}




\section{Introduction}

The study of dynamics in Keplerian potentials is an old yet very progressive area of research. The secular orbital evolution of light (test) particles in the dominating central potential accompanied by a distant perturber is one of the classical problems in celestial mechanics. According to the pioneering works of \citet{Kozai1962} and \citet{Lidov1962}, the orbital solution within this hierarchical three-body setup is often called Kozai--Lidov (K--L) dynamics. Various works have extended its original formulation, which supposed a non-evolving circular orbit of the perturber, e.g., eccentric perturber \citep{naozHotJupitersSecular2011,lithwickECCENTRICKOZAIMECHANISM2011}, relativistic effects \citep{naozRESONANTPOSTNEWTONIANECCENTRICITY2013a,limRelativisticThreebodyEffects2020}, mass loss and transfer \citep{michaelySECULARDYNAMICSHIERARCHICAL2014}.

Considering the four-body setup brings new degrees of freedom and also more variants of the general setup \citep{huangVeryRestrictedFourBody1960,simoRelativeEquilibriumSolutions1978,scheeresRestrictedHillFourBody1998,baltagiannisEquilibriumPointsTheir2011}. A possible configuration has recently been investigated by \citet{haasSecularTheoryOrbital2011a}. Similarly to K-L dynamics, their setup consists of a dominating central body and a massive perturber on a circular orbit. Contrary to K-L dynamics, they considered the orbital evolution of \emph{two} light, mutually gravitationally interacting bodies inner to the orbit of the massive perturber. In their work, \citet{haasSecularTheoryOrbital2011a} focused on the case when the two inner orbits are close to each other in terms of semi-major axes and are initially co-planar (with arbitrary inclination with respect to the orbit of the perturber). An additional assumption, primarily imposed due to limitations of the used calculus, was the non-evolving zero eccentricity of the two inner orbits. \citet{haasSecularTheoryOrbital2011a} argue that this assumption is relevant if another non-Keplerian spherically symmetric potential is present within the system, being strong enough to damp the K-L oscillations of the inner bodies enforced by the massive perturber. Within this setup, \cite{haasSecularTheoryOrbital2011a} developed a secular theory showing that the two inner orbits periodically exchange their angular momentum such that their inclinations oscillate. If their mutual interaction is strong enough (which depends on their mass and separation), the precession of their orbits is synchronised, i.e., the initial co-planar structure is nearly preserved. In the other case, orbital planes of the inner bodies precess differentially due to the perturbing force of the outer body, which leads to disruption of the co-planar configuration. We refer to the temporal evolution of the specific four-body setup introduced by \cite{haasSecularTheoryOrbital2011a} as the \emph{VHS mechanism} throughout this paper.

A shortcoming of the secular theory of VHS dynamics is the requirement of the spherically symmetric external potential needed to dampen the Kozai-Lidov oscillations, which reduces its applicability in observed astrophysical systems. However, \citet{haasCouplingYoungStellar2011,haasSecularTheoryOrbital2011a} introduced a physically realistic setup in which the VHS mechanism is applicable. They studied a system in which the super-massive black hole (SMBH) in the Galaxy's centre, Saggitarius A* (\sgra) \citep{ghez_first_2003,eisenhauer_sinfoni_2005,gillessen_monitoring_2009,gillessen_orbit_2009,yeldaIncreasingScientificReturn2010}, represents the dominant body, and the additional spherical potential is due to the surrounding nuclear star cluster. They considered the perturbing body to be the circum-nuclear gaseous disc \citep{martinSurvivingHoleSpatially2012, liuMILKYWAYSUPERMASSIVE2012, hsiehMolecularGasFeeding2017,tsuboiALMAViewCircumnuclear2018,goicoecheaHighspeedMolecularCloudlets2018,hsiehCircumnuclearDiskRevealed2021} and the bodies on inner nearby co-planar orbits to be the observed stars from the young stellar disc that is within the distances of $0.04$ -- $0.4\,\pc$ from the central super-massive black hole \citep{levin2003stellar,paumardTwoYoungStar2006b,bartkoEvidenceWarpedDisks2009a,bartkoExtremelyTopHeavyInitial2010a}. \citet{haasCouplingYoungStellar2011,haasSecularTheoryOrbital2011a} suggested that the four-body dynamics in the spherically symmetric external potential can explain the specific, near-perpendicular orientation of the stellar disc with respect to the distant perturber.

Our study aims to expand the scope of the VHS dynamics described in \cite{haasSecularTheoryOrbital2011a} and to explore its applicability in a broader range of astrophysical systems by relaxing some of the assumptions of the underlying secular theory. Firstly, we develop the idea, suggested in the original work, that the non-Keplerian spherical potential can be omitted if we consider post-Newtonian terms in the gravity of the central body while still working within the secular approach. Secondly, we investigate the evolution of systems with a non-zero eccentricity of the two inner orbits by directly integrating the equations of motion. Finally, we consider a scenario in which the eccentricity of the inner orbits evolves over time, i.e., when the K-L oscillations are not entirely damped.

The paper is structured as follows: In Section~\ref{sec:setup}, we provide a detailed description of the four-body setup we are studying. Section~\ref{sec:secular_theory_zeroecc} provides a summary of the secular theory developed in \cite{haasSecularTheoryOrbital2011a}, along with a discussion of the damping of K-L oscillations due to the effects of general relativity. Section \ref{sec:results} describes several examples of systems with non-zero eccentricity that were integrated. Finally, we present our conclusions on the generalised VHS dynamics in Section~\ref{sec:conclusions}.

\begin{figure*}
\includegraphics[width=\linewidth]{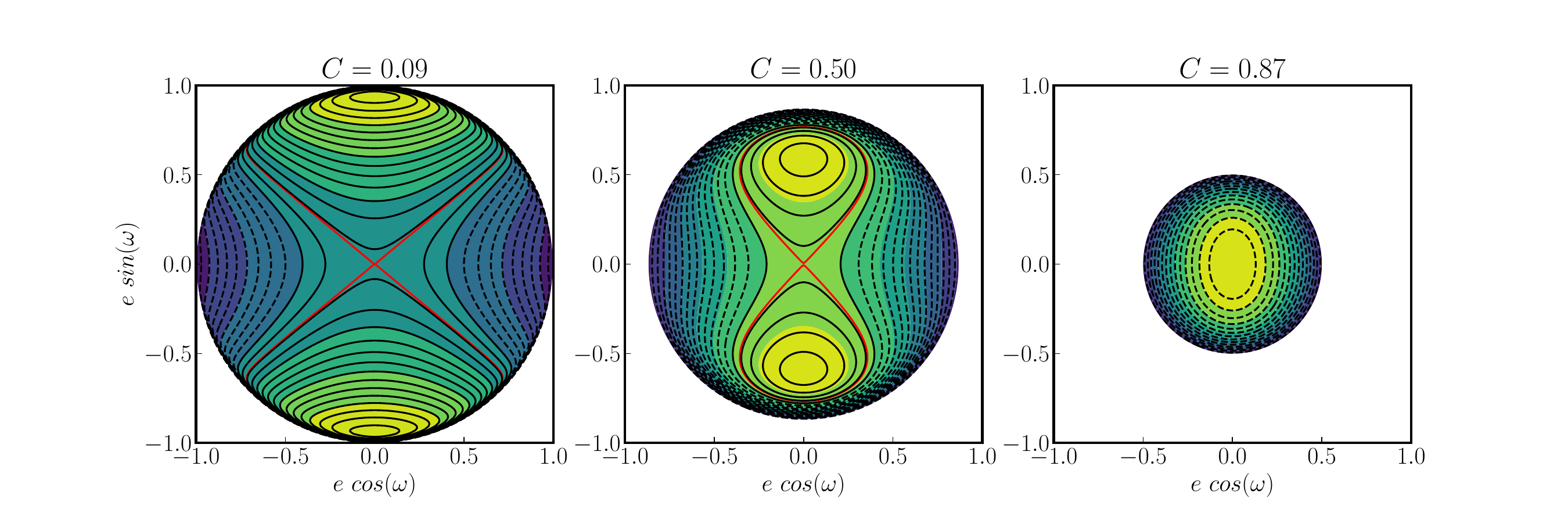}
\caption{Isocontours of $\rbar$ for different fixed values of the Kozai integral ($C$, indicated above individual panels) in a three body system showing pure Kozai--Lidov dynamics. The shape of these isocontours is independent on $a$. The leftmost panel shows isocontours in $e-\omega$ space for $C<\sqrt{3/5}$, which shows large changes in eccentricity over the orbit. The middle panel is for $C$ still smaller than $\sqrt{3/5}$ but with separatrix (displayed with red line) reaching smaller values of eccentricity. In the rightmost panel we have $C>\sqrt{3/5}$ and we see oval evolutionary tracks, meaning small changes in eccentricity. }
\label{fig:kozai_isocontours}
\end{figure*}
\section{Setup}
\label{sec:setup}

We study a hierarchical four-body system with a dominant central body, characterized only by its mass, $M_\bullet$. The system further consists of a distant perturber of mass $M_\mathrm{p}$ on a circular orbit with radius $R_\mathrm{p}$ around the central body. The orbit of the perturber defines the reference plane. We can choose any line within this plane to define our reference axis to calculate the longitude of the ascending node, $\Omega$.
Finally, we consider two light particles of masses $m$ and $m^\prime$ where $m, m^\prime \ll M_\mathrm{p}$ on orbits around the central body with a semi-major axes $a$ and $a^\prime$, which are much smaller than $R_\mathrm{p}$ and $a^\prime<a$. These light bodies are in inclined orbits, having inclinations $i$ and $i^\prime$ with respect to the reference plane. The last important parameters we consider are the longitudes of the ascending nodes of the two bodies, $\Omega$ and $\Omega^\prime$. Initial conditions are set up such that $i=i^\prime$ and $\Omega=\Omega^\prime$. 

As an example, in this work we use the objects observed in the Galactic Centre as an astrophysical system to provide us realistic values for $M_\bullet$, $M_\mathrm{p}$ and $R_\mathrm{p}$. We set the system that may correspond to the situation in the vicinity of the {\sgra} black hole, i.e., $M_\bullet=4 \times 10^6 \mathrm{M_\odot}$ \citep{ghez_first_2003,eisenhauer_sinfoni_2005,gillessen_monitoring_2009,gillessen_orbit_2009,yeldaIncreasingScientificReturn2010}. We consider the distant perturber of mass of $M_\mathrm{p}=10^4 \mathrm{M_\odot}$ and a semi-major axis of $R_\mathrm{p}=0.1$ pc, aiming to mimic the overall gravitational influence of the observed clockwise young stellar disc (CWD) \citep{lu_slar_2013,von_fellenberg_young_2022}. The two light particles could be representatives of the S-stars that are observed in the Galactic Centre.

\section{Secular theory}

\label{sec:secular_theory_zeroecc}
In this Section, we follow the mathematical approach used in \citet{haasSecularTheoryOrbital2011a} and briefly sketch the main ideas. In particular, we consider a secular approach to describe the long-term evolution of the system described in Sec.~\ref{sec:setup}. For this, the mean interaction potential of the system, $\rbar$, averaged over fast changing variables needs to be specified. As it can be given as a direct sum of the individual terms describing different components of the system, we discuss these separately in the following sections.

\subsection{Potential of the distant / outer perturber}
The averaged interaction potential between a perturbing body on a circular orbit with radius $R_\mathrm{p}$ and a particle on an orbit with semi-major axis $a$, eccentricity $e$ and inclination $i$ with respect to the orbital plane of the perturber reads \citep{Kozai1962}:
\begin{equation}
\rbar_\mathrm{p} = - \frac{G m M_\mathrm{p}}{16 R_\mathrm{p}} \left( \frac{a}{R_\mathrm{p}} \right)^2 [(2+3e^2)(3\cos^2{i} -1) + 15e^2 \sin^2{i} \cos{2\omega} ]
\label{eq:Rperturber}
\end{equation}
where $\omega$ is the argument of periapses of the orbit. Suppose $\rbar_\mathrm{p}$ is the only component of the total perturbing potential (i.e., the system is reduced to a three-body setup). In that case, the body on the inner orbit is subject to quadrupole K--L dynamics \citep{Kozai1962,Lidov1962}. Depending on the initial conditions, its eccentricity and inclination may undergo large periodic variations that are mutually coupled through the so-called Kozai integral, $C \equiv \sqrt{1 - e^2}\cos{i}$, which, together with the semi-major axis ($a$) and $\rbar_\mathrm{p}$, is a conserved quantity along the orbit evolution.

The number of known integrals of motion allows for an effective insight into the K--L dynamics through plots of isocontours of $\rbar$ in the $e$-$\omega$ space, which for fixed values of $a$ and $C$ give sets of possible evolutionary tracks (see Figure~\ref{fig:kozai_isocontours}). These sets form two qualitatively different topologies: For $C > \sqrt{3/5}$, they consist of concentric ovals, which means that the eccentricity oscillates slightly along the evolutionary path and $\omega$ rotates within the whole range $(0, \pi)$ (see right panel of Figure~\ref{fig:kozai_isocontours}). If $C \leq \sqrt{3/5}$, the topology qualitatively changes: a separatrix crosses the central point; It divides the diagram into zones with $\omega$ librating around the value of $\pi / 2$ or $3\pi / 2$ and the outer rotation zone (left and middle panel of Figure~\ref{fig:kozai_isocontours}). The lower the value of $C$, the larger the eccentricity oscillations. The characteristic time scale for these oscillations is given by \citep{Kozai1962,Lidov1962}:

\begin{equation}
    T_\mathrm{K} \equiv \frac{M_\bullet}{M_\mathrm{p}} \frac{R_\mathrm{p}^3}{a \sqrt{G M_\bullet a}} .
    \label{KL_timescale}
\end{equation}

An important result from the isocontour plots is that the zero eccentric orbit is stable for $C > \sqrt{3/5}$, while it undergoes periodic variations when $C$ below the limiting value. 

Finally, let us note that the longitude of the ascending node, $\Omega$, rotates monotonically in the full range of $(0, 2\pi)$ for arbitrary initial conditions. The rate of precession depends on the other orbital elements, as well as on the mass and semi-major axis of the perturber. However, the value of $\Omega$ does not affect the evolution of the other orbital elements, which is a natural consequence of the axial symmetry of the problem.

\begin{figure*}
	\centering
    \includegraphics[width=\linewidth]{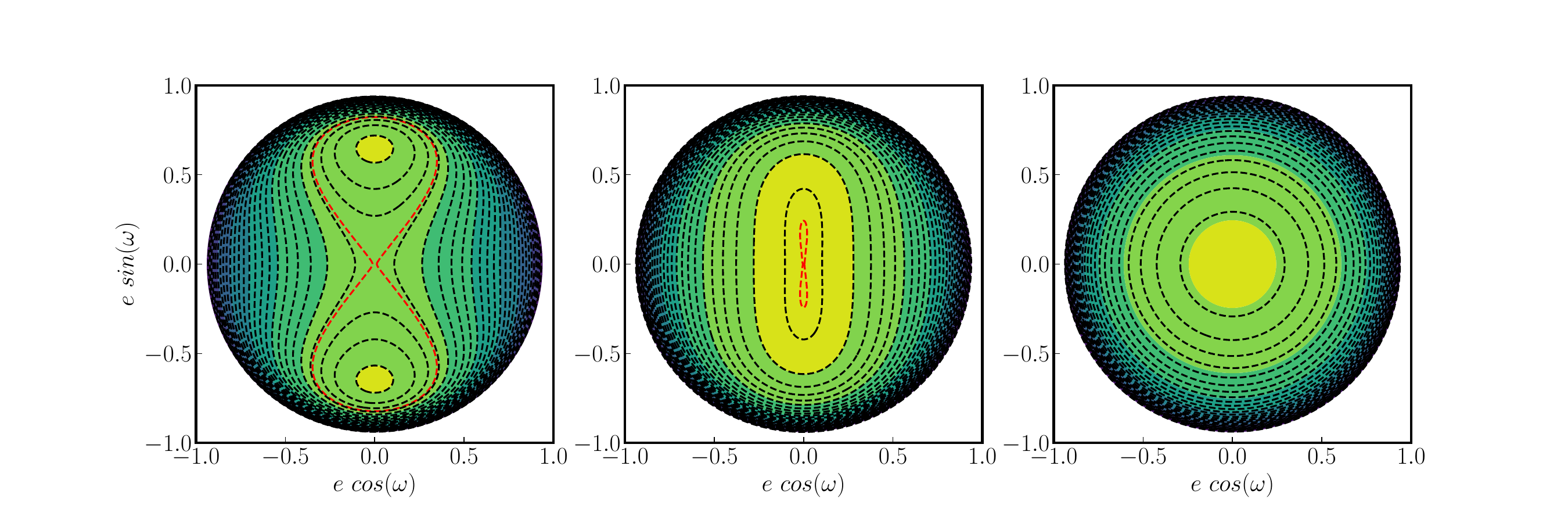}
	\caption{The potential isolines when the potential due to post newtonian approximation is added to the perturbing potential are displayed in three panels, representing two cases with identical parameters except for the semi-major axis of the test particle. In these examples the value of $C=0.34$ which is smaller than the limiting value for K--L dynamics, and the separatrix is shown in red. The panel on the left depicts the instability at $e=0$ when $a=0.2R_{p}$, while the panel on the right illustrates the stability at $e=0$ when $a=0.03R_{p}$. The central panel shows the boundary area when $a=0.145R_{p}$ and the K--L oscillations are damped.}
	\label{fig:stablevunstable}
\end{figure*}
\subsection{Spherical potential}

In our study of four-body systems, we examine two distinct sources of an external spherical potential. The first source is the presence of an extended mass around the central body, while the second source is an approximation of the first-order post-Newtonian corrections to the gravity of $M_\bullet$. Although these sources differ, they have very similar effects and impact the evolution of the two bodies in a similar manner.

\subsubsection{Extended Mass}

\citet{haasSecularTheoryOrbital2011a} and
\citet{haasCouplingYoungStellar2011} considered such an astrophysical context involving an extended mass around the central body, influencing the secular dynamics of the inner orbit(s). In particular, the authors provide an analytic form for the mean potential corresponding to the mass density distribution with power-law profile, $\rho_c \propto r^{\beta-2}$,
\begin{equation}
\rbar_\mathrm{c}= -\frac{GmM_{c}}{\beta R_\mathrm{p}}\left(\frac{a}{R_\mathrm{p}}\right)^\beta \mathcal{J}(e,\beta),
\label{eq:rbar_c}
\end{equation} 
where $M_c$ stands for the integral of the extended mass density within the orbit of the perturber ($R_\mathrm{p}$) and
\begin{equation}
\mathcal{J}(e,\beta)=\frac{1}{\pi}\int_0^\pi (1-e \cos{u})^{1+\beta} \; \text{d}u = 1 + \sum_{n\geq 1}a_n e^{2n}.
\label{eq:J_function}
\end{equation} 
The coefficients $a_n$ are given by 
\begin{equation} 
\frac{a_{n+1}}{a_n}=\left[1- \frac{3+\beta}{2(n+1)}\right]\left[1- \frac{2+\beta}{2(n+1)}\right],
\end{equation}
with $a_1=\beta (1+\beta)/4$.

From the spherical symmetry of this perturbing potential we get that its only manifestation on the orbit evolution is a monotonous (retrograde) rotation of the argument of pericentre, $\omega$.
When combined with the potential of the distant perturber $\rbar_\mathrm{p}$, the potential of the extended mass generally leads to damping of the Kozai--Lidov oscillations (see \citet{HaasSubr2021} for a detailed discussion). This damping stabilizes the zero eccentricity orbit for arbitrary inclination for a suitable choice of system parameters. Note also that in such a situation, monotonous rotation of the longitude of the ascending node remains the primary manifestation of the influence of the distant perturber. \citet{subrWarpedYoungStellar2009} showed that for damped K--L oscillations, the rate of change in longitude of the ascending node is given by 

\begin{equation}
    \frac{\text{d}\Omega}{\text{d} t} \approx - \frac{3}{4} \frac{\cos{i}}{T_\mathrm{K}} \frac{1+\frac{3}{2}e^2}{\sqrt{1-e^2}} \approx \text{constant.}
    \label{eq:dOmegadt}
\end{equation}

This equation shows that when the K--L oscillations are damped, $\frac{\text{d}\Omega}{\text{d} t}$ depends on the semi-major axis through $T_\mathrm{K}$ (see \autoref{KL_timescale}) and will result in differential precession for different orbits.

\subsection{Post-Newtonian corrections}
It has already been discussed in the literature that relativistic corrections to the gravity of the central body can play a role similar to the spherical potential of the extended mass in secular dynamics \citep{holman1997chaotic,blaesKozaiMechanismEvolution2002,karasEnhancedActivityMassive2007}, enforcing a (prograde) rotation of the argument of the pericentre, $\omega$. A straightforward way to implement this relativistic effect within the framework presented above is to use the approximation given by \citet{rubincamGeneralRelativitySatellite1977}. This approximation mimics the rotation of the argument of pericenter due to the relativistic effect of the central body using an additional spherically symmetric potential,
\begin{equation}
V_\mathrm{GR}=-\frac{GM_\bullet h^2}{c^2 r^3},
\label{eq:rubincam_V}
\end{equation} 
where $h \equiv \sqrt{GM_\bullet a(1-e^2)}$ is the specific angular momentum of the test particle and $c$ stands for the speed of light. Formally, this potential is equivalent to spherical mass distribution with density profile $\rho \propto r^{-5}$, that is, the form of the averaged potential given by \autoref{eq:rbar_c} with $\beta = -3$ may be directly used, giving us the mean potential of the first-order post-Newtonian correction,
\begin{equation}
\rbar_\mathrm{GR} = -\frac{GM_\bullet m h^2}{ c^2 a^3} \mathcal{J}(e,-3).
\label{eq:rbar_gr}
\end{equation}
Note that, in comparison to the general mean potential for extended mass distribution, \autoref{eq:rbar_gr} contains additional dependence on eccentricity through $h$, and it has one less parameter ($M_\bullet$ vs. $M_c$ and $a_P$). Also note that \autoref{eq:rbar_gr} diverges as $e$ approaches unity. 


\begin{figure}
	\centering
	\includegraphics[width=\linewidth]{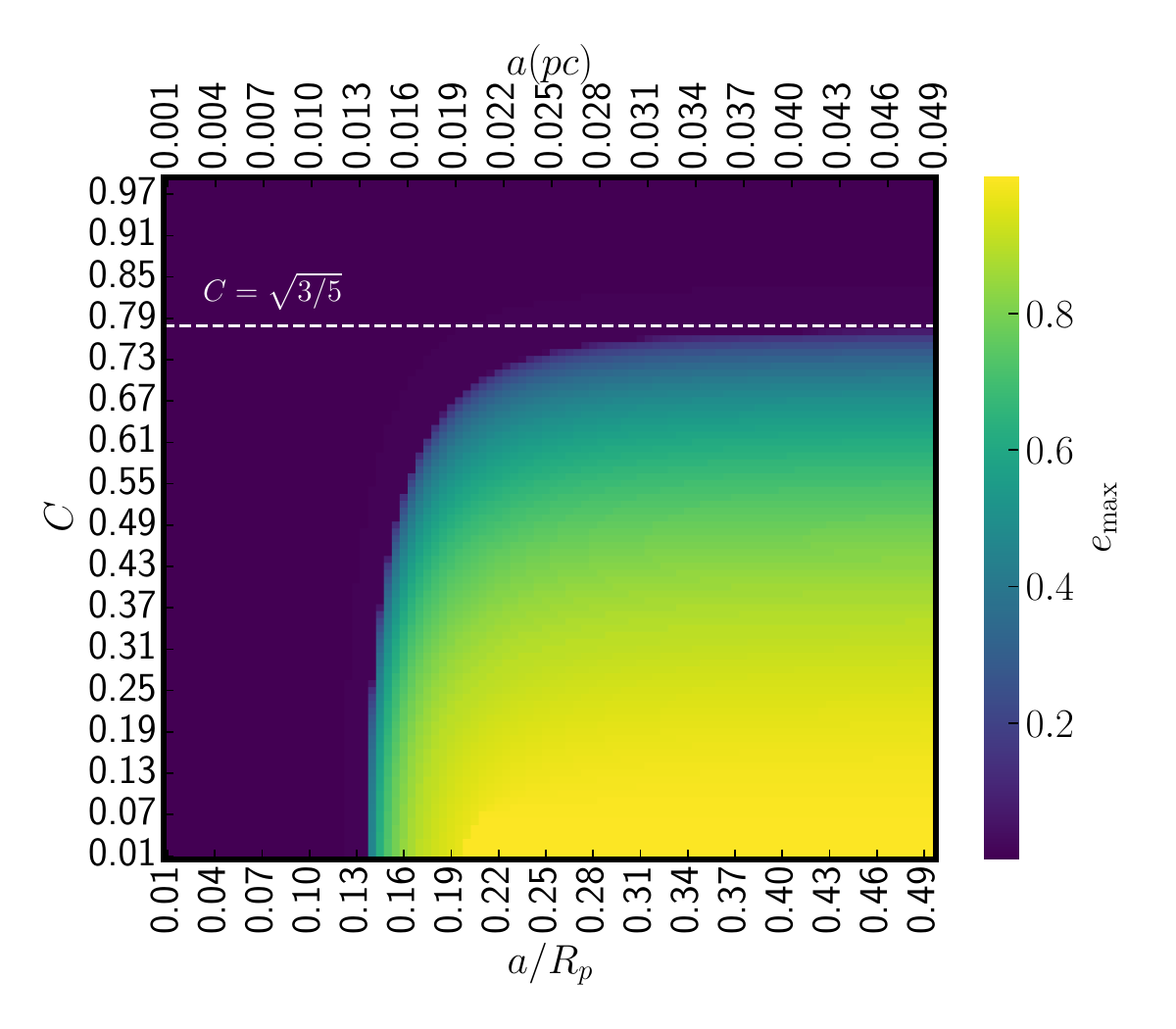}
	\caption{Heatmap of largest variation in eccentricity for a $1 M_{\odot}$ star initially on an orbit of eccentricity $e=10^{-4}$ in the combined potential of a central dominant body ($M_\bullet=4 \times 10^6 \mathrm{M_\odot}$) and a distant perturber ($M_\mathrm{p}=10^4 \mathrm{M_\odot}$ , $R_\mathrm{p}=0.1$ pc) on a circular orbit. The lower value of maximum eccentricity in a significant range of the parameter space shows that the relativistic corrections due to the SMBH partially or entirely dampen the Kozai--Lidov oscillations.}
	\label{fig:GRvCusp}

\end{figure}

We visualise the damping effect of the post-Newtonian corrections using the isocontours of the perturbing potential in the $e$-$\omega$ space in \autoref{fig:stablevunstable}. In particular, we show three examples of $\rbar(e,\omega)$ with $\rbar = \rbar_\mathrm{p} + \rbar_\mathrm{GR}$ for a randomly selected value of $C = 0.34$ and the properties of the central body and perturber are same as \sgra ($M_\bullet=4\times10^6 \mathrm{M_\odot}$) and the CWD ($M_\mathrm{p}=10^4 \mathrm{M_\odot}$, $R_\mathrm{p}=0.1$pc) as described in Section \ref{sec:setup}. We change the value of the semi-major axis of the inner body, i.e., with variable strength of $\rbar_\mathrm{GR}$ with respect to $\rbar_\mathrm{p}$.

In the left panel of \autoref{fig:stablevunstable}, the topology is very similar to that of the middle panel of \autoref{fig:kozai_isocontours}, which means that $\rbar_\mathrm{p}$ dominates over $\rbar_\mathrm{GR}$ in absolute value for most of the parameter space. The middle panel of \autoref{fig:stablevunstable} shows a setup with a smaller value of semi-major axis, leading to a decrease in the absolute value of $\rbar_\mathrm{p}$ while, at the same time, it leads to a growth in the absolute value of $\rbar_\mathrm{GR}$, which means that it contributes considerably to $\rbar$. The topology of the isocontours of $\rbar$ remains the same as in the previous case, but the overall structure changes so that the separatrix does not reach smaller eccentricity values. Further reduction of the semi-major axis, as shown in the right panel of \autoref{fig:stablevunstable}, leads to $\rbar_\mathrm{GR}$ fully dominating over $\rbar_\mathrm{p}$, and hence the isocontours of $\rbar$ form nearly circular shapes as a consequence of the independence of $\rbar_\mathrm{GR}$ on $\omega$. In this case, the K--L oscillations are strongly damped, and the zero eccentricity orbit becomes stable and does not evolve.

For the sake of the analytic treatment of the four-body dynamics described in the following sections, the system configuration must be such that the zero eccentricity orbit is stable. However, due to the non-trivial dependence of $\rbar_\mathrm{GR}$ and $\rbar_\mathrm{p}$ on system parameters, this condition must be evaluated from case to case.

In Figure~\ref{fig:GRvCusp}, we evaluate it for parameters of the system that may correspond to the situation in the vicinity of the {\sgra} black hole and the semi-major axis of the inner body is sampled within the range $0.01 - 0.5 R_\mathrm{p}$, which falls into the region of the S-stars for the example setup described in Sec \ref{sec:setup}. We quantify the damping of K--L oscillations by evaluating the maximum value of eccentricity $e_\mathrm{max}$ reached by the system during its evolution when starting from near-zero eccentricity. The K--L oscillations are successfully damped when we obtain smaller values of $e_\mathrm{max}$,  as the only source of change in eccentricity in these systems is the K--L dynamics. We see that in this setup, the GR effects damp the K--L oscillations for the entire range of $C$ for $a \lesssim 0.14 R_\mathrm{p}$. At the same time, for $a \gtrsim 0.3 R_\mathrm{p}$ the K--L dynamics is less affected; that is, the zero eccentric orbit is stable only for $C\gtrsim\sqrt{3/5}$, shown by the white dashed line in \autoref{fig:GRvCusp}.

\subsection{Inter-particle potential}
\label{sec:inter_particle_potential}
In order to describe the four body setup, \citet{haasSecularTheoryOrbital2011a} evaluated the averaged inter-particle potential for {\em circular} orbits,

\begin{equation}
\rbar_\mathrm{i} = -\frac{Gmm^\prime}{a} \Psi ( \alpha, \boldsymbol{n} \cdot \boldsymbol{n^\prime}),
\label{eq:rbar_i}
\end{equation}
where $\alpha \equiv a^\prime/a$. $\boldsymbol{n}$ and $\boldsymbol{n^\prime}$ are the unit vectors that are normal to the mean orbital plane for the two stars, which can be parameterized as $\boldsymbol{n} = \lbrack\sin{i}\sin{\Omega}, -\sin{i}\cos{\Omega}, \cos{i}\rbrack^T$ and $\boldsymbol{n^\prime}=\lbrack\sin{i^\prime}\sin{\Omega^\prime}, -\sin{i^\prime}\cos{\Omega^\prime}, \cos{i^\prime}\rbrack^T$. We can define the function $\Psi$ as 

\begin{equation}
    \Psi(\zeta, x) = \sum_{l \geqslant 2} [ P_l(0)]^2 \zeta^l P_l(x)\;,
    \label{psi_eq}
\end{equation}
where $P_l(x)$ are the Legendre polynomials. 

We can express the potential energy due to the interaction of inner circular orbits and the outer perturber,
\begin{align}
\rbar_{\mathrm{p},0} &= -\frac{GmM_\mathrm{p}}{R_\mathrm{p}} \Psi(a/R_\mathrm{p}, \cos{i}),\\
\rbar_{\mathrm{P},0}^\prime &= -\frac{Gm^\prime M_\mathrm{p}}{R_\mathrm{p}} \Psi(a^\prime /R_\mathrm{p}, \cos{i^\prime }).
\label{inter_perturb}
\end{align}

\subsection{VHS mechanism}

The total averaged potential of the four-body setup described in Section \ref{sec:setup} is: 
\begin{equation}
\rbar = \rbar_i + \rbar_{\mathrm{p},0} + \rbar_{\mathrm{p},0}^\prime
\label{eq:rbar_vhs}
\end{equation}
and the classical orbital elements are assumed to evolve according to the Lagrange equations

\citep[see, e.g.][]{bertottiPhysicsSolarSystem2003}: 
\begin{align}\label{eq:lagrange1}
    \frac{\text{d} \cos{i}}{\text{d}t}&=-\frac{1}{m\eta a^2}\frac{\partial \overline{\mathcal{R}}}{\partial \Omega}, & \frac{\text{d} \Omega }{\text{d}t}&=-\frac{1}{m\eta a^2}\frac{\partial \overline{\mathcal{R}}}{\partial \cos{i}}, \\
     \frac{\text{d} \cos{i^\prime }}{\text{d}t}&=-\frac{1}{m^\prime \eta^\prime {a^\prime}^{2}}\frac{\partial \overline{\mathcal{R}}}{\partial \Omega^{\prime}},  & \frac{\text{d} \Omega^{\prime} }{\text{d}t}&=-\frac{1}{m^\prime \eta^\prime {a^\prime}^2} \frac{\partial \overline{\mathcal{R}}}{\partial \cos{i^\prime }},
 \label{eq:lagrange2}
\end{align}
here $\eta$ and $\eta^\prime$ are the mean motion frequncies of the two bodies. Although the average potential due to either the extended mass or relativistic corrections plays an essential role in damping the K--L oscillations of the circular orbits, we may omit it here as it does not contribute to the target subset of Lagrange equations, \autoref{eq:lagrange1}) \& \ref{eq:lagrange2}.

The set of \autoref{eq:lagrange1} \& \ref{eq:lagrange2} with mean perturbing Hamiltonian (\autoref{eq:rbar_vhs}) has been first studied by \citet{haasSecularTheoryOrbital2011a} and we refer to their solution in general as the \emph{VHS mechanism}. These equations may be translated to equations for normal vectors, $\boldsymbol{n}$ and $\boldsymbol{n^\prime}$, of the orbital planes \citep[Equations 21-26]{haasSecularTheoryOrbital2011a}.

\begin{equation}
\begin{gathered}
    \frac{\text{d}\boldsymbol{n^\prime }}{\text{d} t} = \omega^\prime_\mathrm{I} \ (\boldsymbol{n^\prime } \times \boldsymbol{n}) + \omega^\prime_\mathrm{p}\ (\boldsymbol{n^\prime } \times \boldsymbol{e}_z),\\ \frac{\text{d}\boldsymbol{n}}{\text{d} t} = \omega_\mathrm{I} \ (\boldsymbol{n} \times \boldsymbol{n^\prime }) + \omega_\mathrm{p}\ (\boldsymbol{n} \times \boldsymbol{e}_z),
\end{gathered}
    \label{dnprimedt}
\end{equation}
where
\begin{align}
    \omega^\prime_\mathrm{I}= -\eta^\prime \alpha \left(\frac{m}{M_\bullet} \right) \Psi_x(\alpha,\boldsymbol{n\cdot n^\prime }), & \ \ \  \omega_\mathrm{I}= -\eta \left(\frac{m^\prime }{M_\bullet} \right) \Psi_x(\alpha,\boldsymbol{n\cdot n^\prime })
    \label{w_I}
\end{align}
\begin{align}
    \omega^\prime_\mathrm{p}= -\eta^\prime \left(\frac{M_\mathrm{p}}{M_\bullet} \right) \Psi_x\left(\frac{a^\prime }{R_\mathrm{p}},\cos{i^\prime }\right), & \ \  \    \omega_\mathrm{p}= -\eta \left(\frac{M_\mathrm{p}}{M_\bullet} \right) \Psi_x\left(\frac{a}{R_\mathrm{p}},\cos{i}\right), 
    \label{w_CND}
\end{align} and  
\begin{equation}
    \Psi_x (\zeta,x) \equiv \frac{\text{d}}{\text{d}x} \Psi (\zeta,x).
\end{equation}
The frequencies $\omega_\mathrm{I}$, $\omega^\prime_\mathrm{I}$ and $\omega_\mathrm{p}$, $\omega^\prime_\mathrm{p}$ correspond to the frequencies caused by the mutual interaction of the two bodies and the perturber, respectively.

\citet{haasSecularTheoryOrbital2011a} have shown both by means of analysis of the averaged Hamiltonian as well as direct integration of the Lagrange equations that there exist two qualitatively distinct classes of solutions. On a qualitative level, if the masses of the inner orbits are small enough, or their separation (in terms of semi-major axes) is sufficiently large or a combination of both, we call the regime of interaction \emph{weak}. In the opposite case, we call the interaction \emph{strong}. Explicit formula defining boundary between the two modes is not known, nevertheless, an estimate for particular setup can be obtained comparing the frequencies $\omega_\mathrm{I}$ and $\omega^\prime_\mathrm{I}$ to $\omega_\mathrm{p}$ and $\omega^\prime_\mathrm{p}$. In the \emph{strong} mode, $\omega_\mathrm{I},\ \omega^\prime_\mathrm{I} >> \omega_\mathrm{p},\ \omega^\prime_\mathrm{p}$ which means that evolution of the orbital planes described by \autoref{dnprimedt} is governed by the mutual interaction of the inner orbits. On the other hand, if $\omega_\mathrm{I},\,\omega^\prime_\mathrm{I} << \omega_\mathrm{p} , \ \omega^\prime_\mathrm{p}$ the \emph{weak} mode of the VHS mechanism takes place in which the two planes precess differentially due to the gravitational influence of the outer orbit. Note that none of the frequencies $\omega_\mathrm{I}$, $\omega^\prime_\mathrm{I}$, $\omega_\mathrm{p}$ and $\omega^\prime_\mathrm{p}$ are constant over time. Hence, determination of which mode realises cannot be reliably determined from their initial values.


\subsubsection{Weak mode of the inter-particle interaction}
\label{sec:weak_int_zero}
In the \emph{weak} mode of the VHS mechanism, the two orbits periodically interchange their angular momenta such that their magnitudes stay constant, but mutual inclination changes. The longitudes of their ascending nodes rotate at different rates while still being mutually influenced. This independent rotation of $\Omega$ disrupts the original co-planar configuration. At the moments when $\Delta \Omega \equiv \Omega^\prime - \Omega$ reaches the value of a multiple of $2\pi$, the relative inclination of the two orbits drops to zero, and the planar structure is re-established for the moment. 

\begin{figure}
	\centering	
	\includegraphics[width=\linewidth]{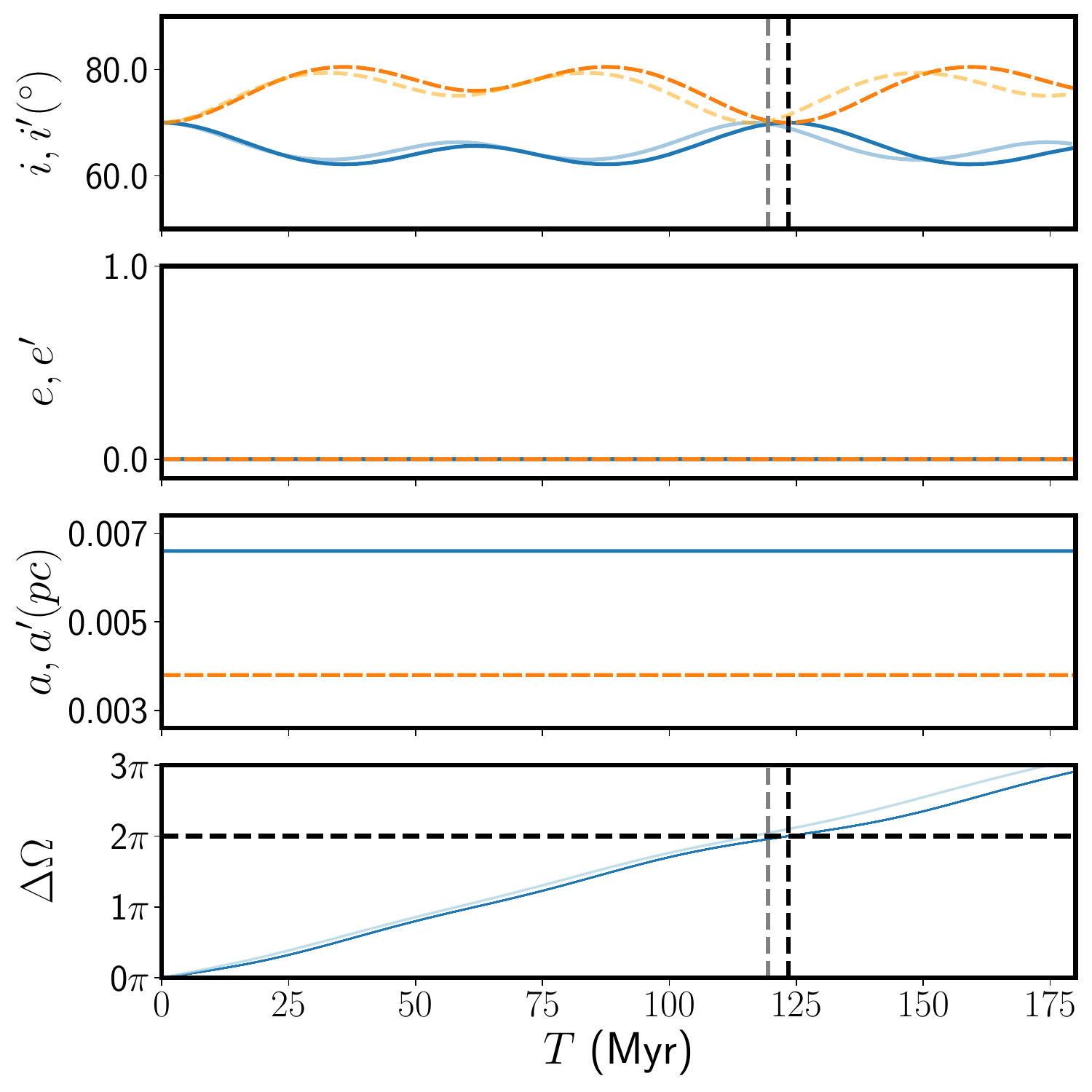}
	\caption{Evolution of model \model{1} showing the weak mode of evolution in VHS mechanism. Weak mode of VHS mechanism results in different rate of change in evolution of $\Delta \Omega$ along with oscillations in inclination. The lighter version of the lines is the result of integration of secular equations while the darker versions are the result of integration of equations of motion. The black dashed lines highlight that the mutual inclination becomes 0 when the value of $\Delta \Omega$ is a multiple of $2\pi$ in the integration of equations of motion. Similarly the grey line is for the integration of secular equations.}
	\label{fig:lowmass_e0_weak}
\end{figure}

An example of this solution is shown in \autoref{fig:lowmass_e0_weak} with parameters of the system given in \autoref{tab:int_parameters} under the label \model{1}. Besides showing the orbital evolution according to the secular approximation, we also plot the evolution of the orbital elements coming from direct integration of the equations of motion. For the latter case, we utilize the {\arwv} integrator \citep{chassonneryARWVCodeUser2019} since it allows for integrations of a few-body system with up to $2.5$ order post-Newtonian approximation. Both solutions share the same qualitative properties with slight differences in the amplitude and period of oscillations of the inclinations which indicate the quality of the secular approximation in this particular configuration. 

Estimate of characteristic time-scale, $T_{\text{char}}$, of the weak mode of the VHS mechanism comes from that (i) the precession of $\Omega$ is dominated by the distant perturber, i.e., it is nearly constant but different for the two inner bodies and (ii) the period of oscillation of inclinations is determined by the time instances  when  $\Omega-\Omega^\prime =2\pi$. To find $T_{\text{char}}$ for which $\Omega(T_\text{char})-\Omega^\prime (T_\text{char})=2\pi$, we can apply \autoref{eq:dOmegadt} independently to both the inner and outer orbits, approximating inclinations and eccentricities with their initial values ($I=I^\prime =I_0$ and $e=e^\prime =e_0$) which yields:

\begin{equation}
    T_\text{char}\approx\frac{16\pi\sqrt{1-e_0^2}}{3(3e_0^2 +2) \cos{I_0}}\Bigg[\frac{1}{T_\text{K}} - \frac{1}{T^\prime _\text{K}} \Bigg]^{-1}.
    \label{eq:char_t}
\end{equation}

For the case of $e_0=0$, this simplifies to the formula given by \citet[Equation 34]{haasSecularTheoryOrbital2011a}. Plugging the inital conditions of \model{1} in \autoref{eq:char_t} we get $T_\text{char}\approx 192$ Myr which is same order of magnitude of $T_\model{1}=123.42$ Myr.

\subsubsection{Strong mode of the inter-particle interaction}
\label{sec:strong_zero}
The \emph{strong} mode of the VHS mechanism occurs when the masses of the inner orbits are large and/or their orbits are closer to each other. In this case, inter-particle interaction surpasses the differential precession of $\Omega$ and $\Omega^\prime$ induced by the distant perturber, and the orbits co-rotate. Similarly to the weak case, the two inner orbits keep the magnitude of total angular momenta constant, yet exchange angular momentum so that their inclinations undergo mirrored oscillations. The amplitude of these oscillations is typically smaller than in the weak mode and, therefore, the two orbits stay nearly coplanar during the whole course of the secular orbit evolution.

\begin{figure}
	\centering	
 	\includegraphics[width=\linewidth]{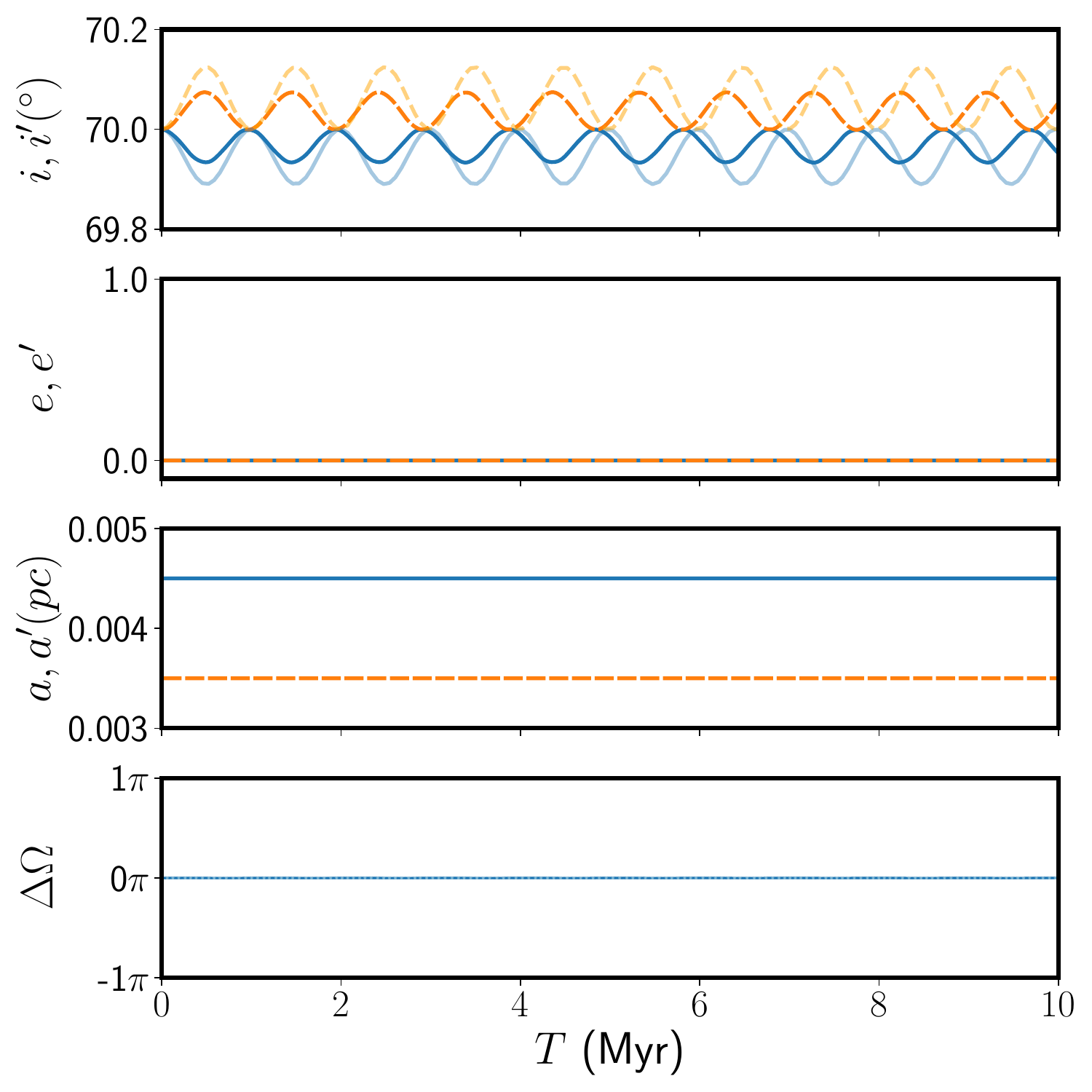}
 	
	\caption{These graphs show the evolution of the orbital parameter of model \model{3}. The stronger mutual interaction between the two stars results in strong mode of VHS mechanism, resulting in a constant $\Delta \Omega = 0^\circ$, which is a complete overlap for integration of both secular and equations of motion. There are tiny inclination oscillations for both integration of equations of motion (darker) and secular equations (lighter), which have slightly different amplitude and period. }
	\label{fig:lowmass_e0}
\end{figure}

\autoref{fig:lowmass_e0} shows a typical example of \emph{strong mode} which occurs in the setup with initial conditions labelled as model \model{3} in \autoref{tab:int_parameters}. Just like the weak case, we show results of the orbital evolution according to the secular approximation as well as direct integration of the equations of motion. The solutions are qualitatively the same which proves that the secular theory is suitable for understanding the nature of the VHS mechanism.

As the differential precession of $\Omega$ is suppressed in this mode of the VHS mechanism, it cannot be used to define any characteristic time-scale. Instead, the fact that $\omega_\mathrm{I}$ and $\omega^\prime_\mathrm{I}$ dominate in \autoref{dnprimedt} can be used to estimate period of the orbital evolution as $T_\text{char} = 2 / (\omega_\mathrm{I} + \omega^\prime_\mathrm{I})$. For \model{3} we get $T_\text{char} \approx 0.98$ Myr, while the numerical integrations give a period of $ 0.97$ Myr.

\section{Numerical Solutions}

\label{sec:results}

The secular theory discussed above relies on the assumption of constant zero eccentricity of the two nearby orbits. This section aims to investigate the four-body dynamics that relax this strict constraint. Since no analytic theory is formulated for the non-zero eccentricity case (and is expected to be non-trivial as the dynamics of nearby eccentric orbits is susceptible to chaotic behaviour induced by close encounters), we study our desired setups with sufficient accuracy using direct numerical integrations. 

The lack of analytic theory makes it difficult to define distinct classes of possible evolution. Our strategy is then to perform a set of integratons with different initial conditions and compare the results with the ideal cases (zero eccentricity) for which we have an analytic insight. 
Therefore, the set of examples presented below is likely to be incomplete in terms of all the possible outcomes but it still shows that the two basic modes of the VHS mechanism have identifiable effects in more general setups.

\autoref{tab:int_parameters} lists the initial conditions of the four setups we discuss in this section, along with the zero eccentricity cases discussed in the previous section. A large number of direct integrations with relativistic corrections using {\arwv} were conducted; We selected a subset of the runs to clearly demonstrate the strong and weak modes of the VHS mechanism when we relax certain requirements for secular theory. These individual cases are discussed separately in the following sections.

\begin{table}
\begin{center}
 \begin{tabular}{l|c|c|c|c|c|}
  \hline
  Model & m,m' & a' & a & e,e' & Figure\\
   & ($\mathrm{M_\odot}$) & ($pc$) & ($pc$)\\
  \hline
      \model{1} & 1 & $0.0035$ & $0.007$ & 0.0 & \autoref{fig:lowmass_e0_weak}\\
    \model{2} & 1 & $0.0035$ & $0.007$ & 0.721 & \autoref{fig:rand_ecc_run_weak}\\
      \model{3} & 10 & $0.0035$ & $0.0045$ & 0.0 & \autoref{fig:lowmass_e0}\\
  \model{4} & 10 & $0.0035$ & $0.0045$ & 0.77 & \autoref{fig:rand_ecc_run}\\

  \model{5} & 10 & $0.0196$ & $0.0215$ & 0.01 & \autoref{fig:unstable_strong}\\
  \model{6} & 10 & $0.0151$ & $0.0168$ & 0.03 & \autoref{fig:unstable_weak}\\

  \hline
 \end{tabular}
\end{center}
  \caption{Parameters of the two light bodies in the four-body setup. For all the models the parameters of the central body and the perturber stay consistant. The central dominant body is at the origin and has mass $M_\bullet=4\times 10^6 \mathrm{M_\odot}$. The perturber has a mass of $M_\mathrm{p}=10^4 \mathrm{M_\odot}$ in a circular orbit at radius $R_\mathrm{p}=0.1$ pc.}
\label{tab:int_parameters}
\end{table}

\subsection{Weak mode with \textit{e} > 0}

\begin{figure}
	\centering
	\includegraphics[width=\linewidth]{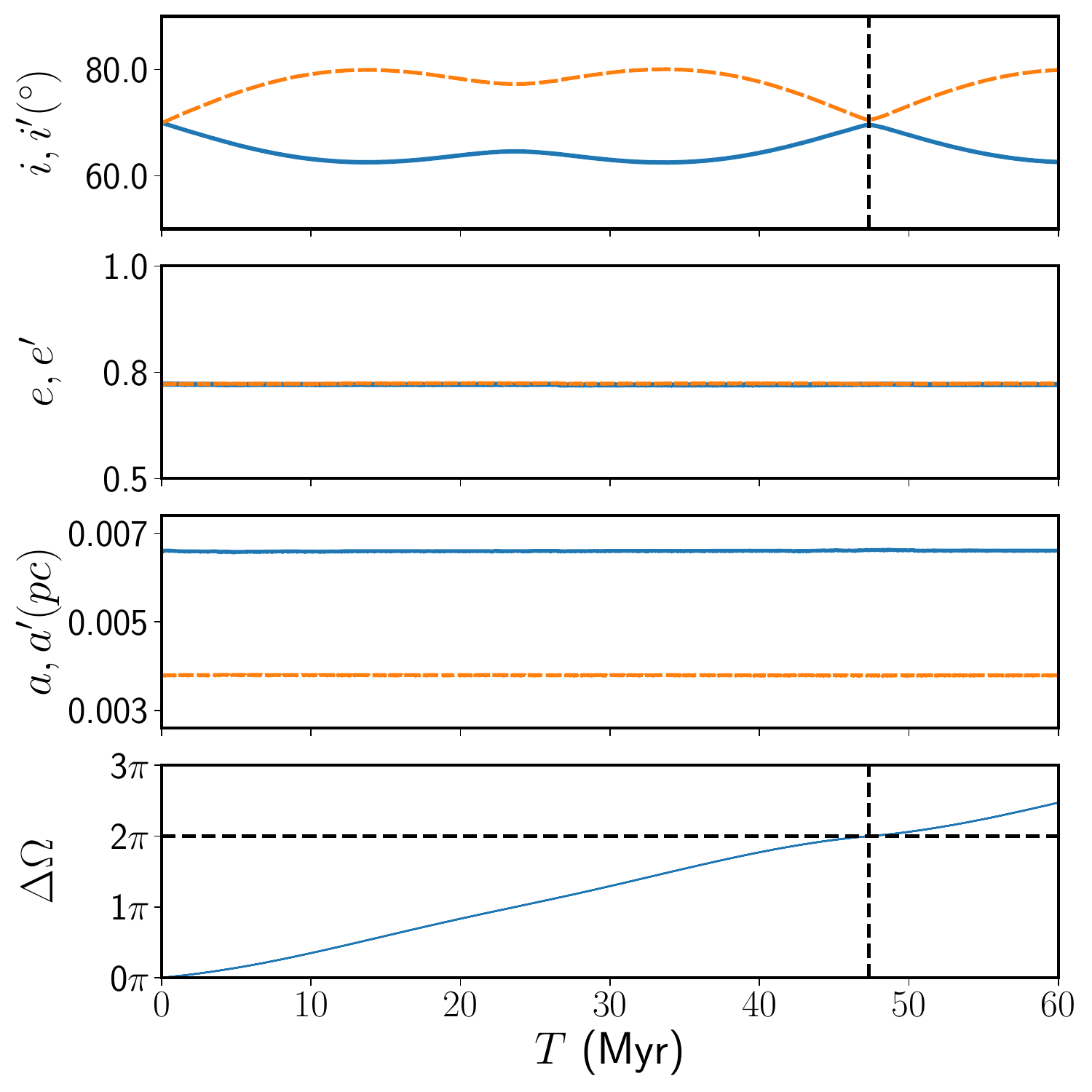}

	\caption{Example of weak mode of VHS mechanism with non-zero eccentricity. It shows the evolution of the model \model{2} which is similar to \model{1} but with non-zero eccentricity. The evolution is similar to \autoref{fig:lowmass_e0_weak} and we again see $\Delta i =0$ when $\Delta \Omega = 2\pi$ as shown by the black dashed lines.}

	\label{fig:rand_ecc_run_weak}
\end{figure}

Let us start by relaxing the condition of zero eccentricity of the two nearby orbits. The model \model{2} is then straightforwardly derived from \model{1} simply by changing the initial eccentricity values from zero to $0.721$. Since the K--L oscillations are damped, the eccentricity of the orbits does not evolve. We can see this in the temporal evolution of selected orbital elements of the two particles for this setup in \autoref{fig:rand_ecc_run_weak}. When $\Omega -\Omega^\prime$ reaches a multiple of $2 \pi$, the relative inclination of the two particles drops to zero. This directly agrees with the angular momentum exchange in the weak mode of VHS mechanism between the two bodies as described in Sec. \ref{sec:weak_int_zero}.

 Period of the secular evolution within the weak mode of the VHS mechanism for model \model{2} is clearly shorter with respect to the circular case (\model{1}). This is in accord with the dependence of \autoref{eq:char_t} for characteristic time-scale on eccentricity. For model \model{2} it gives $T_\text{char}\approx 75$ Myr, while the period determined directly from the numerical integrations is $T_\model{2} \approx 47$ Myr.

\subsection{Strong mode with \textit{e} > 0}
\label{sec:non-zero-strong}

\begin{figure}
	\centering
	\includegraphics[width=\linewidth]{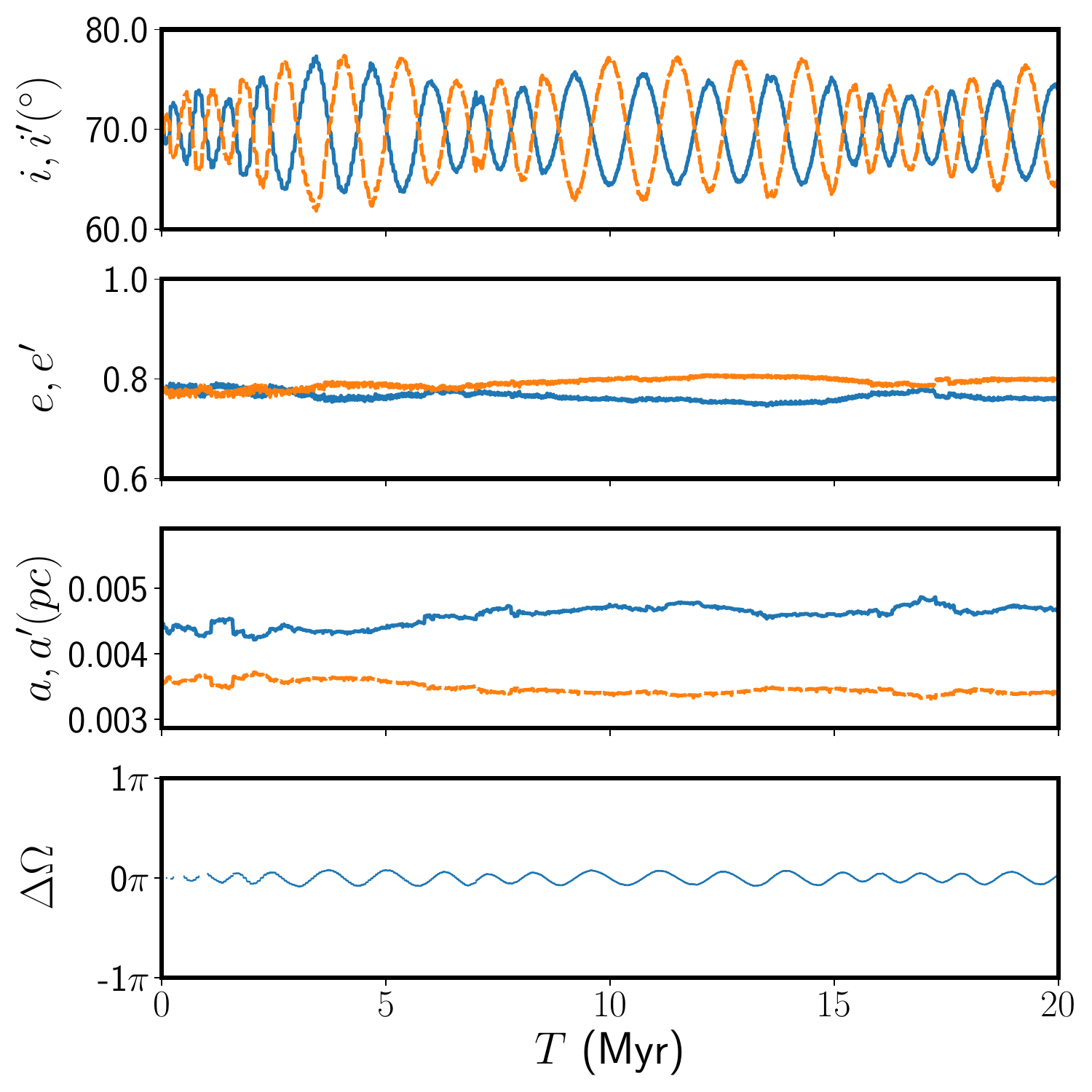}

	\caption{This figure exemplifies how the strong mode of VHS mechanism behaves with non-zero eccentricity. It show the evolution of model \model{4} which is similar to \model{3} but with orbits with eccentricity of $e=0.77$.}

	\label{fig:rand_ecc_run}
\end{figure}

In another example, we consider a system based on \model{3}, but with an initial eccentricity of 0.77 and we refer to this model as \model{4}. \autoref{fig:rand_ecc_run} shows the evolution of the orbital elements of this model. We observe that the inclinations of the two bodies exhibit mirrored oscillations, while the value of $\Delta \Omega$ oscillates around zero. These two signatures suggest that the system is influenced by the strong mode of VHS mechanism, albeit with some qualitative differences compared to the zero eccentricity case.

Contrary to the previous cases (\model{1} -- \model{3}), the orbits undergo non-periodic changes of their semi-major axes, which means that there is a stochastic energy exchange occurring between the two particles. We attribute this to particles on two nearby \emph{eccentric} orbits occasionally getting so close to each other that the instantaneous two-body scattering noticeably affects their semi-major axes and eccentricities. These scattering events mean that we cannot treat the orbital evolution as secular. 

A clear distinction between \model{3} and \model{4} is the evolution of the inclination of the two particles. In \model{3}, the orbits evolve in accordance with the secular theory of \citet{haasSecularTheoryOrbital2011a}, which implies that the inner of the two coplanar orbits is pushed to higher values of inclination while the inclination of the outer orbit decreases. The evolution is more complex in \model{4} compared to \model{3}. In \model{4}, the value of $\Delta i \equiv i^\prime - i$ periodically changes its sign (see Appendix  \ref{appendix:inc_switch} for further discussion). On the other hand, the (quasi)periodic \emph{mirrored} oscillations of inclinations of the two orbits suggest that the angular momentum transfer between them is secular. The magnitude of the change in inclination is also higher in \model{4} compared to \model{3}, but still smaller compared to the inclination oscillations present in weak mode of VHS mechanism (models \model{1} and \model{2}).

Finally, let us focus on the evolution of the longitudes of the ascending nodes $\Omega$ and $\Omega^\prime$ of the two particles. If these were test particles, i.e., not interacting with each other, $\Omega$ and $\Omega^\prime$ would evolve at different constant rates according to \autoref{eq:dOmegadt}, which means that $\Delta\Omega$ would grow monotonically in time, reaching a value of $\approx 28\degr$ on the time scale of $20\,\myr$ in the setup of model \model{4}. However, in the bottom panel of \autoref{fig:rand_ecc_run}, we see limited oscillations of $\Delta\Omega$ around zero with maximum amplitude $\approx 10\degr$. Small $\Delta \Omega$ means that differential precession is suppressed, although not as ideal as in model \model{3} with zero eccentricity.

Considering the two necessary signatures in the evolution of the orbital elements, i.e., small amplitude mirrored oscillations of inclinations and suppressed differential precession in terms of $\Delta\Omega$, we state that the system described in model \model{4} undergoes a generalised mode of the strong mode of VHS mechanism with non-zero eccentricity.

\subsection{Strong mode on the top of Kozai--Lidov cycles}
\label{sec:unstable_strong_kl}

\begin{figure*}
	\centering
    \includegraphics[width=\linewidth]{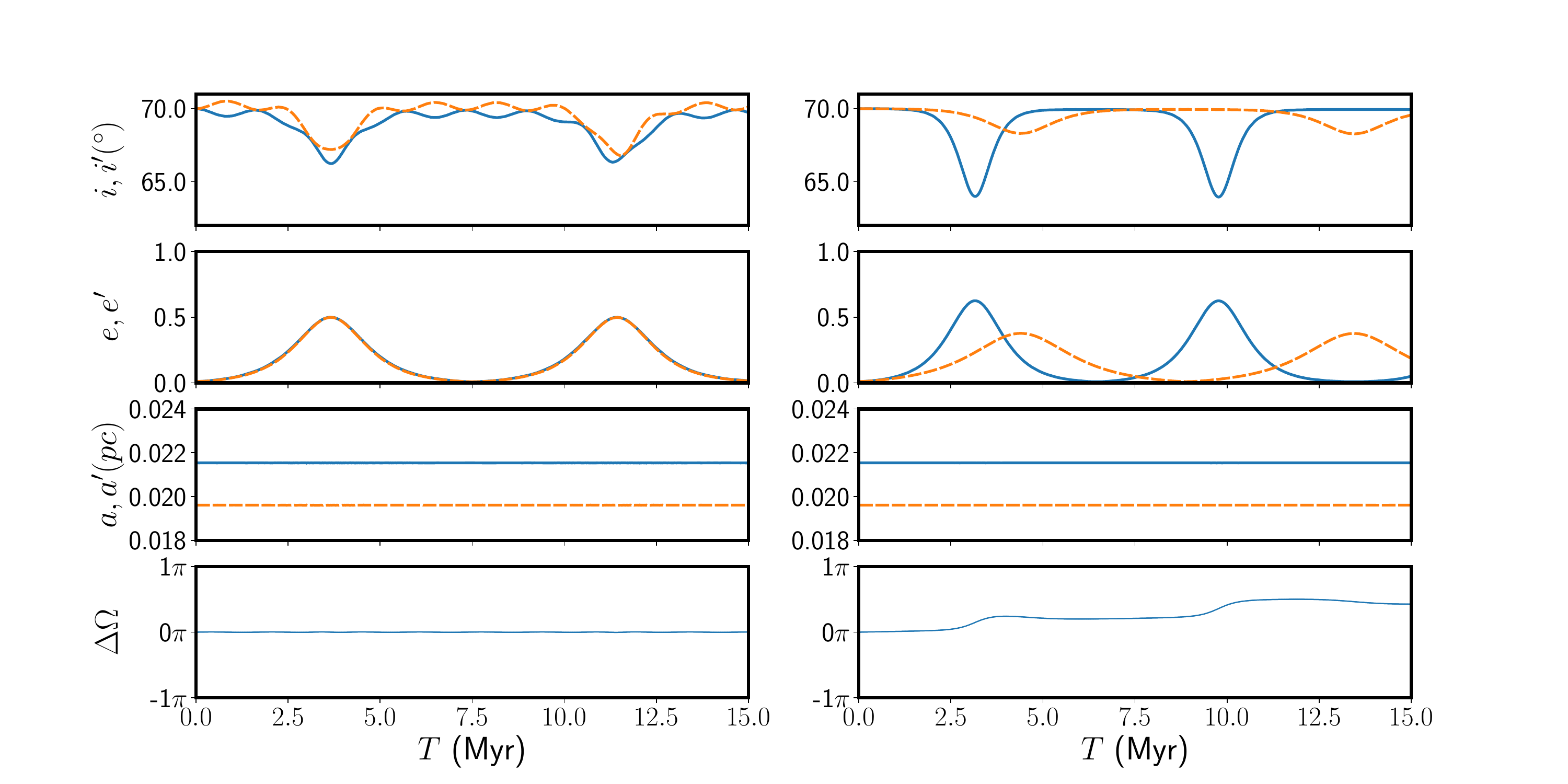}
	
	\caption{The left panel shows an example run of strong mode of VHS mechanism with dynamically evolving eccentricity due to K--L oscillations in model \model{5}. The interaction between the two stars results in a combination of the strong mode of the VHS mechanism and K--L oscillations in both bodies. The strong modes constant $\Delta \Omega$ is still present and the characteristic oscillations in the inclination overlap with the K--L oscillations. However, the right panel shows the evolution of orbital elements when we remove the effects of VHS mechanism by decreasing masses of the two inner particles. The constant zero $\Delta \Omega$ changes to a systematic growth while the two particles have independent K--L oscillations in inclination and eccentricity. }
	\label{fig:unstable_strong}
\end{figure*}
Now that we have seen examples of systems with nonzero eccentricity showing either the weak or strong mode of VHS mechanism, we now try to relax the requirement of having constant eccentricity by reducing the damping of K--L dynamics. We can do this by increasing the ratio of $a/R_\mathrm{p}$, which strengthens the perturbing potential due to the outer body with respect to the damping potential due to the post-Newtonian corrections. We study model \model{5} (\autoref{tab:int_parameters}) to explore the VHS mechanism with variable eccentricity.

The left panels of \autoref{fig:unstable_strong} show the evolution of orbital elements for this system, with the eccentricity oscillations of the two particles now sharing a common period and amplitude. Their inclinations have a more complex evolution, but it is straightforward to identify short-term mirrored oscillations around the mean value. The mean value of the inclination oscillates due to K--L dynamics, which is on a much longer time scale than the strong mode of VHS mechanism. In this case, the inclinations evolve according to the secular theory of \citet{haasSecularTheoryOrbital2011a} in that the inclination of the inner body is always greater than that of the outer one. Finally, it is the suppressed differential precession of $\Omega\ $ and $\Omega^\prime$ which indicates that we see the two particles moving in the regime where strong mode of VHS mechanism is present, i.e. with a mutually locked orientation of their orbital planes while undergoing typical long-term K--L cycles.

Since the particles undergo two independent types of secular evolution at once, we find it beneficial to demonstrate how the orbits will evolve without the VHS mechanism. We can achieve this in the test-particle regime, that is, when mutual interaction between the two inner bodies is suppressed, as shown in the right panels of \autoref{fig:unstable_strong}. Both particles undergo independent K--L oscillations in the test-particle regime with different periods and amplitudes. Difference of the longitudes of the ascending nodes, $\Delta\Omega$, systematically (though not monotonically) grows over time. We can also see the period of the K--L oscillations are different between the left and right figures. This means that the VHS mechanism changes $T_\mathrm{K}$ of the two bodies so that the new $T_\mathrm{K}$ is between the $T_\mathrm{K}$ of the two bodies if they were evolving independently.

Let us also point out the apparent regular nature of this setup contrary to the above-discussed model \model{4} in Section \ref{sec:non-zero-strong}. This property, however, is not generic as the system is chaotic; slightly modified initial parameters of the system may lead to dramatically different evolution of orbits.

\subsection{Transition from the strong to weak mode}

\begin{figure*}
	\centering
	\includegraphics[width=\linewidth]{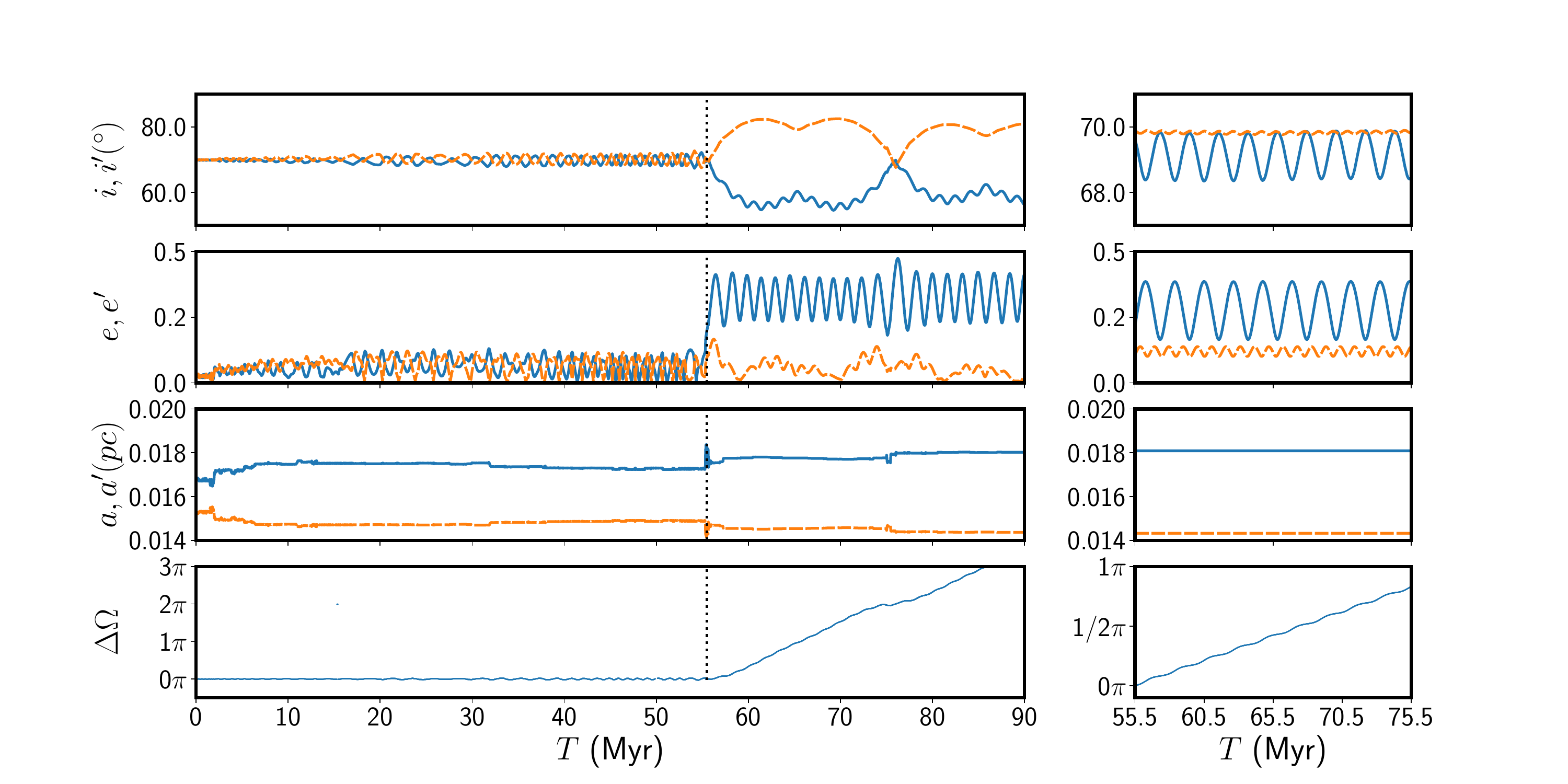}

	\caption{The left panel shows an example run of weak mode of VHS mechanism with dynamically evolving eccentricity (model \model{6}). The initial strong mode of VHS mechanism between the two stars is changed due to stochastic effects and results in the stars separating. This leads to a combination of weak mode of VHS mechanism and the K--L oscillations in the blue body. The characteristic oscillations in the inclination are present but overlap the K--L oscillations in the blue body. In the right figure we show how the system would have evolved after the timestep ($T=55.5$ Myr) marked by the black dashed line if there had been no mutual interaction between the two particles, and thus no VHS mechanism. We see that the orbits have a consistant K--L oscillations without any extra oscillations in inclination.}
 \label{fig:unstable_weak}
\end{figure*}

It has been demonstrated already in Section \ref{sec:non-zero-strong} (model \model{4}) that the systems with non-zero eccentricity may be subject to slightly chaotic evolution due to stochastic close encounters between the two inner particles. Model \model{6} in \autoref{tab:int_parameters} is another example of a system where such encounters play an essential role. One notable difference from \model{4} is that the initial eccentricity in the current setup is close to zero but not precisely zero. The left panels of \autoref{fig:unstable_weak} show the temporal evolution of the setup \model{6}.

From the beginning, until $T \approx 56\,\myr$, it shows an evolutionary pattern similar to that of model \model{4}, i.e., inclinations of the two inner particles undergo mirrored oscillations with $\Delta i$ periodically changing its sign. At the same time, $\Delta \Omega$ oscillates around zero value, meaning the two orbits co-rotate and are almost co-planar, i.e., the orbits undergo the strong mode of VHS mechanism. Also similar to model \model{4} is the stochastic (though rather subtle) evolution of semi-major axes and eccentricities.

At $T\approx 56\,\myr$, another close encounter of the two inner particles leads to a more substantial perturbation of their orbits in semi-major axes and eccentricities. Subsequent evolution shows that this event led to the transition from the weak to the strong mode: the inclinations of the two particles exhibit larger amplitude mirrored oscillations. At the same time, longitudes of the ascending nodes precess differentially. At the moments when $\Delta\Omega$ reaches a natural multiple of $2\pi$, both orbits share the same value of inclination, i.e., they are co-planar for that short period.

Another remarkable feature during the phase of weak mode is
short-periodic oscillations of eccentricity and inclination of the outer particle. These are K–L oscillations induced by the outer perturbing body that now become less damped because of a suitable angular momentum and energy change. To confirm the nature of these oscillations, we show the evolution of a system of two test particles in the external potential with initial conditions taken from the state of M6 shortly after the two-body scattering event at $T = 55.5$ Myr in the right panels of \autoref{fig:unstable_weak}. These lighter bodies then have the following orbital parameters: $ a=0.0183pc,\ a^\prime = 0.0146pc,\ e=0.21,\ e^\prime = 0.11,$ and $\Delta i= 0.202^\circ$. The outer particle, which is more influenced by the distant perturber, undergoes coupled regular oscillations of eccentricity and inclination. In contrast, the oscillations of the inner particle are strongly damped due to the stronger effect of the relativistic precession.

\subsection{Disc like structures}
\label{sec:disc}

\begin{figure}
\includegraphics[width=\linewidth]{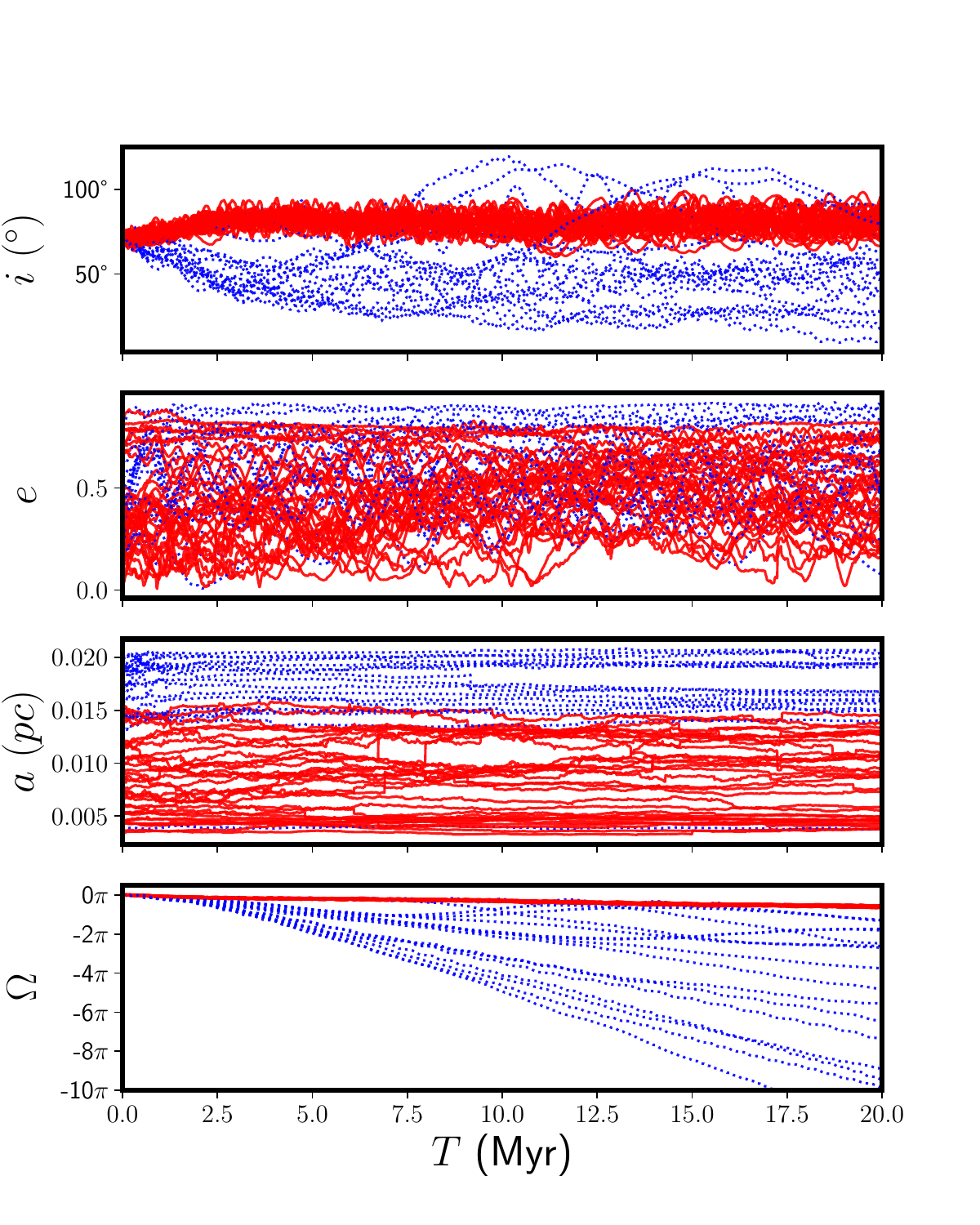} 
\caption{Evolution of orbital elements of individual stars within the model described in Section \ref{sec:disc}. Evolutionary tracks depicted with red solid lines correspond to orbits that are part of the coherent structure the whole integration time. Blue dotted lines correspond to orbits that get more separated from the coherent structure for at least some period of time.}
\label{fig:subim1}
\end{figure}

\begin{figure}
\includegraphics[width=\linewidth]{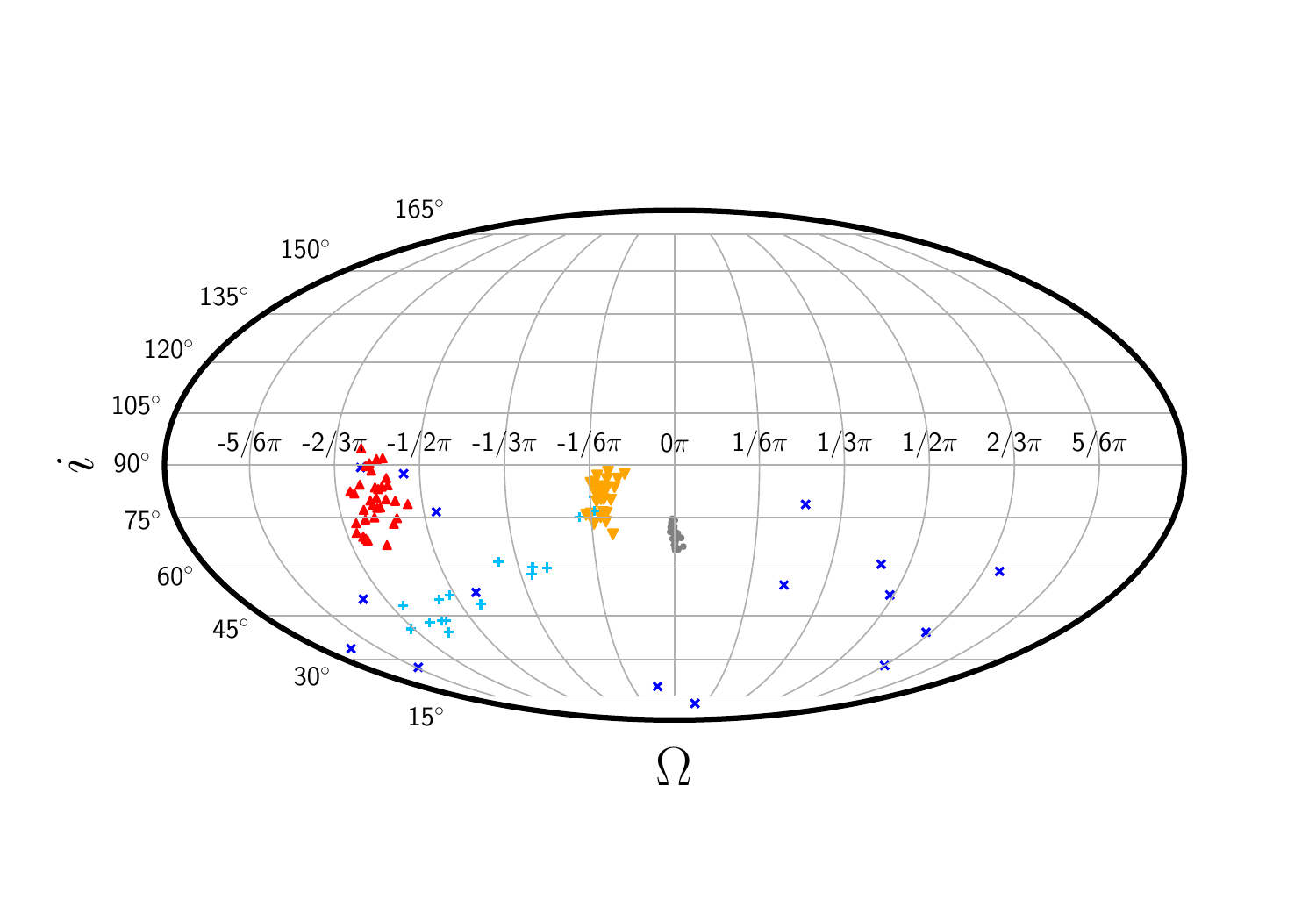}
\caption{Projection of angular momentum vectors of individual orbits of the system discussed in Section \ref{sec:disc}. Red (triangle) and blue points (X) represent final states ($T=20\,\myr$) with the colour coding being the same as in \autoref{fig:subim1}. Orange (upside down triangle) and light blue points (+) represent the state of the red and blue orbits at $T=2.5$ Myr, respectively. The grey points shows the initial state of the system.}
\label{fig:subim2}
\end{figure}

Let us demonstrate the VHS mechanism in the evolution of a $N$-body system. We study a setup inspired by \citet{haasCouplingYoungStellar2011} but with two significant differences. First, the initial eccentricities of the orbits are uniformly distributed within the range $[0,1)$, while \citet{haasCouplingYoungStellar2011} initially considered circular orbits. Second, the post-Newtonian corrections to the central body's gravity dampen the K--L oscillations instead of the extended mass distribution included in \citet{haasCouplingYoungStellar2011}. 

We consider a hypothetical disk of 50 stars orbiting around \sgra, a supermassive black hole of mass $M_\bullet=4\times10^6 \mathrm{M_\odot}$. The disk is perturbed by a massive perturber of mass $M_\mathrm{p} = 1\times10^4 \mathrm{M_\odot}$ orbiting \sgra on a circular orbit at $R_\mathrm{p}=0.1$ pc. The masses of the stars in the disk are sampled from a Salpeter distribution function, $\xi(m) \propto m^{-2.35}$, in the mass range $1-15 \mathrm{M_\odot}$. 

For all orbits, the initial values of the argument of pericentre $\omega$ and the longitude of the ascending node $\Omega$ are set to zero. At the same time, other orbital elements are sampled uniformly with $a \in [0.0035,0.02)\,\pc$, $e \in [0.0,1.0)$, $i \in [65^\circ, 75^\circ)$, and the true anomaly $\nu \in [0, 2\pi)$. We integrate this setup with the same integration code, {\arwv}, which we used in the previous sections.
 
\autoref{fig:subim1} illustrates the temporal evolution of the orbital elements of all 50 stellar orbits. \autoref{fig:subim2}  shows the projection of the normal vectors of the orbital planes of the same orbits at $T=0,\ 2.5\ \text{and}\ 20$ Myrs. We currently separate the stars whose $\Omega$ stays within $20\degr$ of the median of the whole sample throughout the course of evolution and mark them in red in both figures. We refer to this as the disc-like structure as the orbits are co-rotating with each other. The stars depicted in blue are objects whose orbits rotate independently and visually occupy a more spread out region in \autoref{fig:subim2}.

Although the current configuration differs from the model presented in \citet{haasCouplingYoungStellar2011}, the main dynamical effects are qualitatively similar: approximately 2/3 of the orbits, predominantly from the inner region of the disc, maintain the disc-like configuration, characterized by similar values of both $i$ and $\Omega$, throughout the entire course of evolution. The remaining outer orbits precess differentially in terms of $\Omega$, resulting in a scattered structure. However, this structure still exhibits a specific feature, as the inclinations of these orbits are typically smaller than their initial values.

In contrast, the inclination of the coherent structure grows with respect to the initial value, becoming nearly perpendicular to the orbital plane of the outer perturbing body. We interpret this evolution similarly as was done in \citet{haasSecularTheoryOrbital2011a}. Specifically, we suggest that the inner orbits mutually interact in the strong VHS mode. Furthermore, the inner and outer parts of the disc initially act as two bodies that mutually interact in the weak VHS mode. After some time, the outer body loses initial coherency due to the differential precession of the orbits of its individual members, which suppresses the weak mode of VHS mechanism between the inner and outer bodies of the disc.

It is worth noting that the model presented here is scaled so that the coherent structure has spatial dimensions similar to those of the system of S-stars observed in the Galactic Centre. While this paper cannot provide any insight into the role of the VHS mechanism in the dynamical evolution of stars in the Galactic Centre, recent research by \citet{aliKinematicStructureGalactic2020} suggests that coherent disc-like structures can be identified within the S-star cluster. This presents an opportunity to observe the potential effects of the VHS mechanism on the stars in the Galactic Centre.

\section{Conclusions}
\label{sec:conclusions}

In this work, we built on the previous study conducted by \citet{haasSecularTheoryOrbital2011a} that explored the dynamical evolution of two nearby, Keplerian, and initially co-planar orbits under the influence of a massive, distant perturber. The secular theory proposed in \citet{haasSecularTheoryOrbital2011a} assumes constant zero eccentricity of orbits of all bodies (the two inner objects close to the dominant body and the distant perturber). This assumption is only applicable to systems where an additional non-Keplerian spherically symmetric potential is not only present, but is also strong enough to damp K--L oscillations of the two inner bodies caused by the gravity of the distant perturber. The secular theory provides two qualitatively different solutions of orbital evolution of the inner bodies, which we refer to as the weak and strong modes of the VHS mechanism.

Generally, the weak mode applies when the masses of the bodies on the inner orbits are small, and/or their separation in semi-major axes is large. This mode results in independent rotations of the longitudes of the ascending nodes of the two orbits due to the influence of the distant perturber. Additionally, the two orbits periodically exchange their angular momentum, leading to periodic coupled oscillations of their inclinations. However, when $\Delta\Omega$ is an integer multiple of $2\pi$, both inner orbits become co-planar again.

For systems with more massive bodies and/or minor separations between the two inner orbits, the strong mode applies. In this mode, the inner orbits have a common rotation rate of $\Omega$, accompanied by oscillations of small-amplitude inclinations.

This paper demonstrates that the qualitative features of the two modes of VHS mechanism are identifiable in systems where some of the critical assumptions of the secular theory are relaxed. Instead of the external potential of some extended mass distribution, post-Newtonian corrections to the dominant body's gravity can dampen the K--L oscillations. This damping is well understood within the original secular theory of \citet{haasSecularTheoryOrbital2011a} with the first-order post-Newtonian approximation given by \citet{rubincamGeneralRelativitySatellite1977}. By relaxing the need for the extended mass to dampen the K--L oscillations, VHS mechanism applies to a broader range of astrophysical systems, such as compact planetary systems or the innermost regions of galactic nuclei.

We have further studied systems with non-zero eccentricity of the inner orbits. We cannot use the secular theory of \citet{haasSecularTheoryOrbital2011a} to study such a setup. Nevertheless, by directly integrating the equations of motion, we have identified key features of both the weak and strong modes of the VHS mechanism. The main difference we found in these setups compared to the zero eccentricity case is within the strong mode. In this mode, the orbital inclinations of the inner particles may swap, meaning that in some setups, they oscillate around the common starting value. Nonetheless, this does not change the general statement that the orbits co-rotate ($\Delta \Omega \approx 0$) within this evolutionary mode.

In order to achieve a more general setup, we have partially relaxed the assumption of constant eccentricity, which assumes complete damping of K--L oscillations of the inner orbits due to the gravity of the outer perturber. We have presented examples of systems where we observe only partially damped K--L oscillations of the inner orbits.\footnote{It is important to note that we considered post-Newtonian dynamics in all the examples, which means that some level of damping of K--L oscillations due to the relativistic pericentre advance was always present.} The typical features of the VHS mechanism's weak or strong modes are identifiable in these systems.

Finally, we have demonstrated, similarly to \citet{haasCouplingYoungStellar2011} and \citet{haasSecularTheoryOrbital2011a}, that the VHS mechanism applies to more complex systems with a larger set of initially co-planar bodies in a relativistic potential. Recent research by \citet{aliKinematicStructureGalactic2020} suggests that coherent disc-like structures can be identified within the S-star cluster. This opens avenues for observing the possible effects of VHS mechanism in the stars in the Galactic Centre.

In summary, the analytical expression of VHS mechanism described in \citet{haasSecularTheoryOrbital2011a} appears to be a robust phenomenon that can even govern the evolution of systems that do not meet the assumptions of the analytic theory. We have demonstrated through several examples that the VHS mechanism patterns can be found even in systems where instantaneous close encounters significantly affect the orbital evolution. Specifically, the persistent near co-rotating configuration within the strong mode may have straightforward, observationally detectable consequences for a broad range of astrophysical systems, such as compact planetary systems or stellar structures in the innermost regions of galactic nuclei. However, it is essential to note that the strong mode of the VHS mechanism does not \emph{create} co-planar and co-rotating structures within our current understanding; instead, it allows for the survival of existing such structures for extended periods. The weak mode may lead to a specific evolution of its orientation, as shown in Section \ref{sec:disc}, which was discussed for a particular setup in \citet{haasCouplingYoungStellar2011}. 

The result of a more general understanding of the VHS mechanism is a potential application in the Galactic Centre to orbits of the S-star cluster. A consequence of evolving eccentric orbits is the introduction of chaos in these systems, which needs to be understood better. Studying this in more detail can facilitate a deeper understanding of the evolution of disk-like structures with the VHS mechanism. These studies will lead to significant insights into the behaviour of astrophysical systems and contribute to a better understanding of the underlying mechanisms that govern their evolution.

\section*{Acknowledgements}
We thank Sai Sasank Chava and Yugantar Prakash for feedback on the manuscript. We thank David Vokrouhlický for his input on using the Rubincam approximation. MS is supported by the Grant Agency of Charles University under the grant number 179123. L\v{S} and JH acknowledge support from the Grant Agency of the Czech Republic under the grant 20-21855S.

\section*{Data Availability Statement}
The data and tools used to produce the plots in this paper will be shared on reasonable request to the corresponding author.

\bibliographystyle{mnras}
\bibliography{ex} 

\appendix

\section{Inclination Crossing in eccentric strong mode}
\label{appendix:inc_switch}

In Section \ref{sec:non-zero-strong}, we describe qualitative difference of the strong mode of the VHS mechanism with eccentric orbits in comparison to the circular case. It has been argued in \citet{haasSecularTheoryOrbital2011a} that, starting from co-planar configuration, the inclination of the inner orbit always grows, while that of the outer one decreases. An important piece of their argument is that precession of the outer orbit due to the distant perturber is always faster which leads to positive value of $\sin(\Omega^\prime-\Omega)$ which  implicitly occurs in \autoref{eq:lagrange1} and \ref{eq:lagrange2} through the dependence of $\rbar_\mathrm{i}$ on $\boldsymbol{n}.\boldsymbol{n}^\prime$.

\begin{figure}
	\centering	
 	\includegraphics[width=\linewidth]{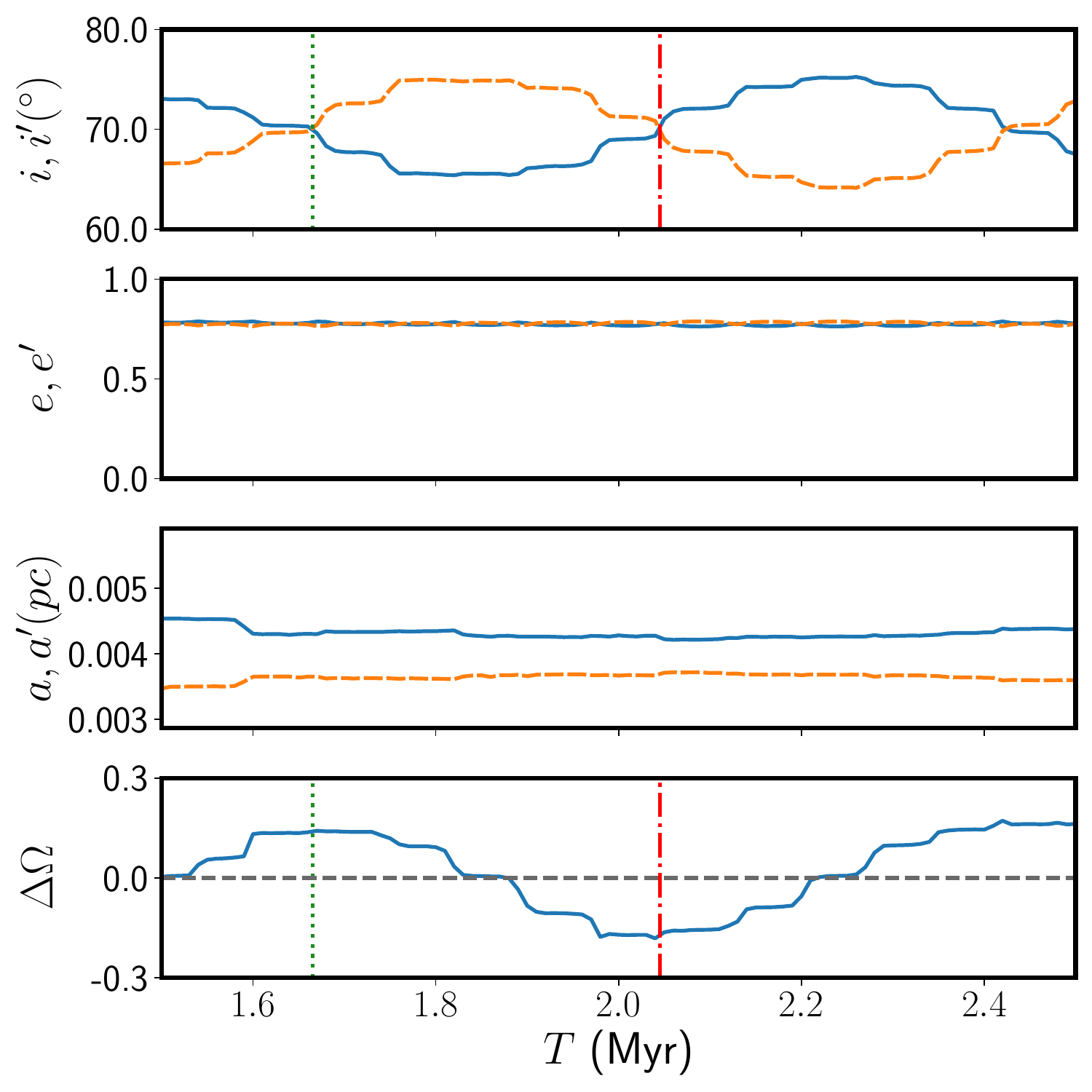}
 	
	\caption{Evolution of the orbital elements within model \model{4} for a short period of time. The two points where $\Delta i=0$ are marked with green (dotted) and red (dash-dot) vertical lines, with appropriate markings in $\Delta \Omega$.}
	\label{fig:high_e_app}
\end{figure}

\begin{figure}
	\centering	
 	\includegraphics[width=\linewidth]{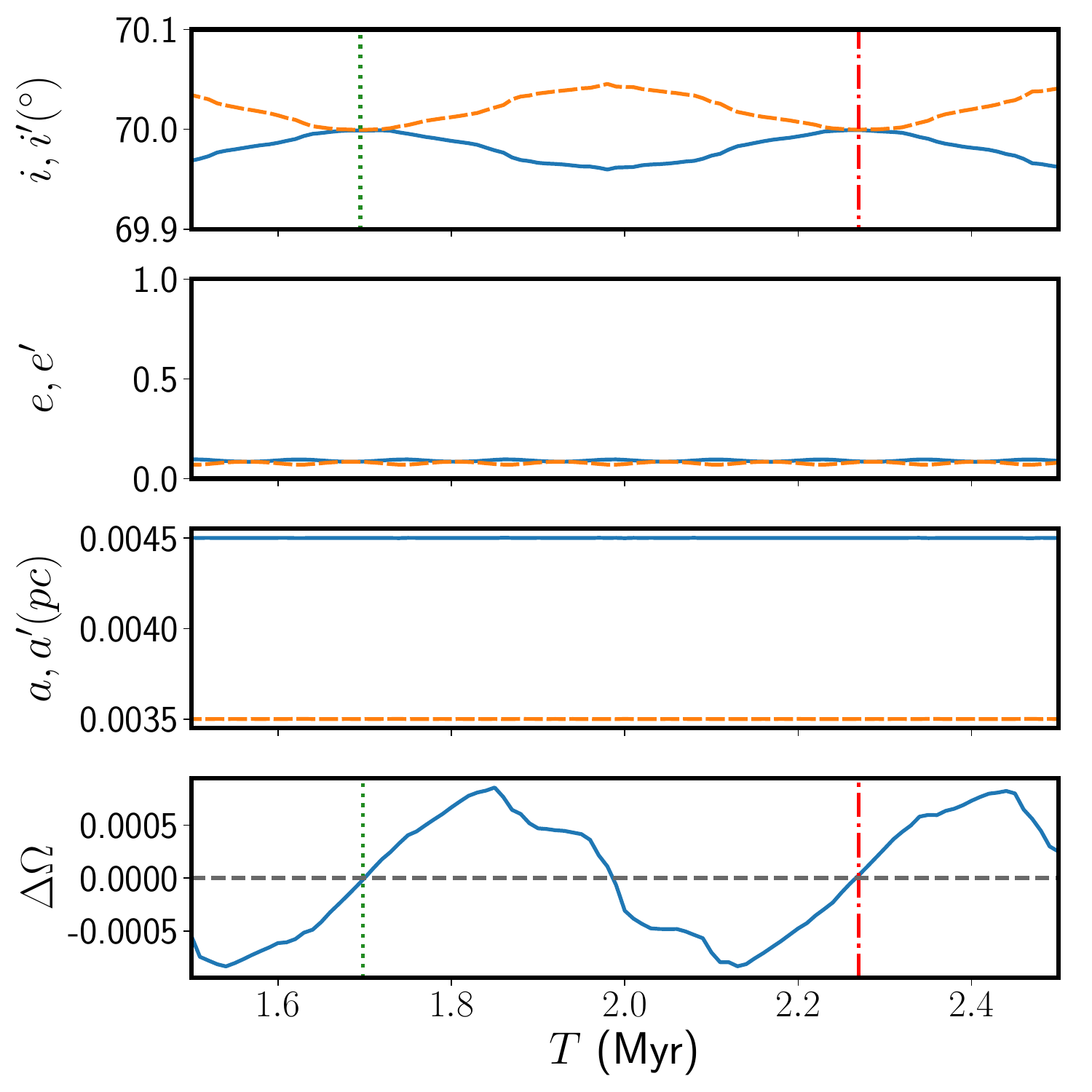}
 	
	\caption{Evolution of the orbital elements within a model similar to model \model{4} but low eccentricity, $e_0 = e^\prime_0 = 0.08$. Green and red vertical lines indicate instances of $i = i^\prime$.}
	\label{fig:low_e_app}
\end{figure}

We don't have secular equations for the VHS mechanism with eccentric orbits in hands, still, we may assume that dependence of $\text{d}i / \text{d}t$ and $\text{d}i^\prime / \text{d}t$ on $\Delta\Omega \equiv \Omega^\prime - \Omega$ is similar to the circular case. \autoref{fig:high_e_app} shows zoomed-in evolution of model \model{4} for a short period of time. Indeed, we see that, in contrary to the circular case, $\Delta\Omega$ reaches non-zero (both positive and negative) vales at the instances of $i = i^\prime$. Depending on the sign of $\Delta\Omega$, inclination of the inner orbits either grows similarly to the circular case ($\Delta\Omega > 0$) or decreases. For comparison, we also show detailed view of evolution of orbital elements for setup similar to model \model{4}, but now with small eccentricities of the two inner orbits, $e_0 = e_0^\prime = 0.08$, in \autoref{fig:low_e_app}. The oscillatory pattern of $\Delta\Omega$ is preserved, but now with (i) several orders of magnitude smaller amplitude and (ii) near zero value at the instances of $i \approx i^\prime$ and (iii) positive derivative at those instances. Evolution of $i$ and $i^\prime$ is then in accord with the analytic argumentation for the zero eccentricity case.

Let us note that due to lack of analytic secular theory for the non-zero eccentricity case, it is hard to discriminate, whether evolution of $\Delta\Omega$ in the strong mode of the VHS mechanism is primarily due to non-uniform precession of the orbits in the field of the distant perturber, or whether it is mainly governed by their mutual torques.

\bsp	
\label{lastpage}
\end{document}